%% file: visguides-2019.tex
\documentclass[journal]{vgtc}                

\ifpdf
  \pdfoutput=1\relax                   
  \usepackage{graphicx}                
  \DeclareGraphicsExtensions{.pdf,.png,.jpg,.jpeg} 
\else
  \usepackage{graphicx}                
  \DeclareGraphicsExtensions{.eps}     
\fi%

\graphicspath{{figures/}{pictures/}{images/}{./}} 

\usepackage{microtype}                 
\PassOptionsToPackage{warn}{textcomp}  
\usepackage{textcomp}                  
\usepackage{mathptmx}                  
\usepackage{times}                     
\usepackage{cite}                      
\usepackage{tabu}                      
\usepackage{booktabs}                  
\usepackage{xcolor}
\usepackage{url}
\usepackage{amsfonts}
\usepackage{amsmath}
\usepackage{amssymb}
\usepackage{enumitem}
\usepackage{flushend}

\onlineid{0}

\vgtccategory{Research}
\vgtcpapertype{Theory/Model}

\title{Studying Visualization Guidelines According to Grounded Theory}

\author{%
Alexandra Diehl, \textit{Member, IEEE},
Matthias Kraus, \textit{Member, IEEE},
Alfie Abdul-Rahman, \textit{Member, IEEE},\\
Mennatallah El-Assady, \textit{Member, IEEE},
Benjamin Bach, 
Robert S. Laramee, \textit{Member, IEEE}, \\
Daniel A. Keim, \textit{Member, IEEE},
Min Chen, \textit{Member, IEEE}}
\authorfooter{
\item%
Alexandra Diehl, Matthias Kraus, Menna El-Assady, and Daniel Keim are with University of Konstanz, Germany. E-mails: \{Alexandra.Diehl, Matthias.Kraus, Mennatallah.El-assady, Keim\}@uni-konstanz.de
\item%
Alfie Abdul-Rahman is with King's College London. UK;
Benjamin Bach is with Edinburgh University, UK;
Robert S. Laramee is with Swansea University. UK;
Min Chen is with University of Oxford, UK.\\
E-mails: alfie.abdulrahman@kcl.ac.uk, 
bbach@inf.ed.ac.uk,
R.S.Laramee@swansea.ac.uk,
min.chen@oerc.ox.ac.uk
}

\shortauthortitle{Diehl \MakeLowercase{\textit{et al.}}:Studying Visualization Guidelines According to the Grounded Theory}

\abstract{%
Visualization guidelines, if defined properly, are invaluable to both practical applications and the theoretical foundation of visualization.
In this paper, we present a collection of research activities for studying visualization guidelines according to Grounded Theory (GT). 
We used the discourses at VisGuides, which is an online discussion forum for visualization guidelines, as the main data source for enabling data-driven research processes as advocated by the grounded theory methodology.
We devised a categorization scheme focusing on observing how visualization guidelines were featured in different threads and posts at VisGuides, and coded all 248 posts between September 27, 2017 (when VisGuides was first launched) and March 13, 2019. To complement manual categorization and coding, we used text analysis and visualization to help reveal patterns that may have been overlooked by the manual effort and summary statistics.
To facilitate theoretical sampling and negative case analysis, we completed an in-depth analysis of the 148 posts (with both questions and replies) related to a student assignment from a visualization course.
Inspired by two discussion threads at VisGuides, we conducted two controlled empirical studies to collect further data to validate specific visualization guidelines.
Through these activities guided by grounded theory, we have obtained a number of new findings about visualization guidelines.
} 


\teaser{
  \centering
  \begin{tabular}{@{}c@{\hspace{4mm}}c@{\hspace{4mm}}c@{}}
    \includegraphics[height=35mm]{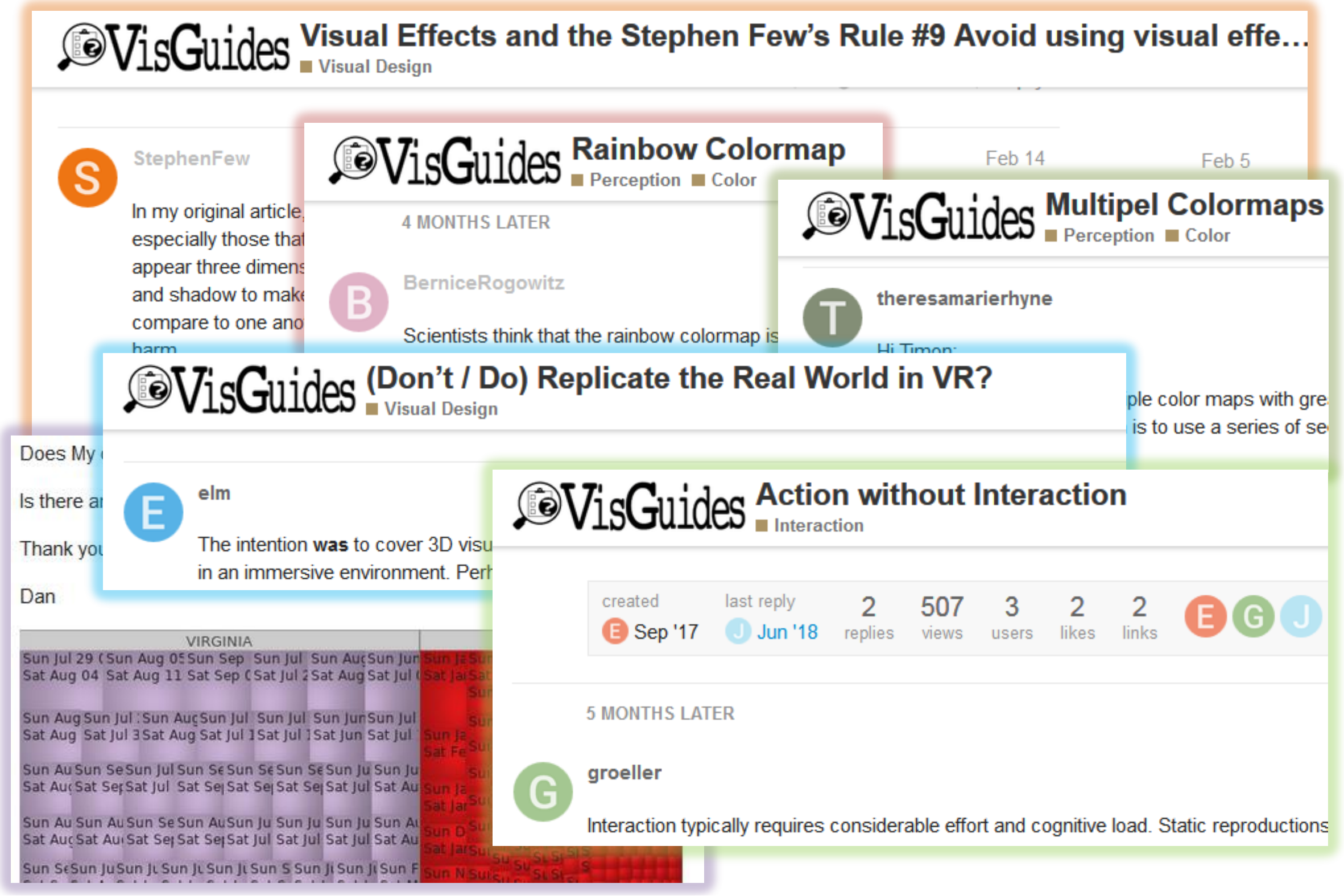} &
    \includegraphics[height=35mm]{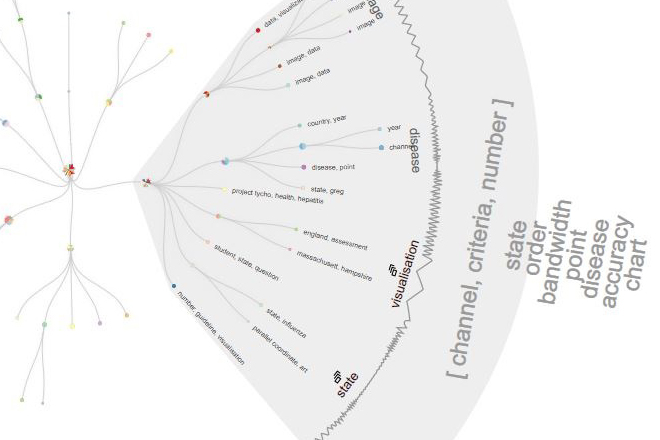} &
    \includegraphics[height=35mm]{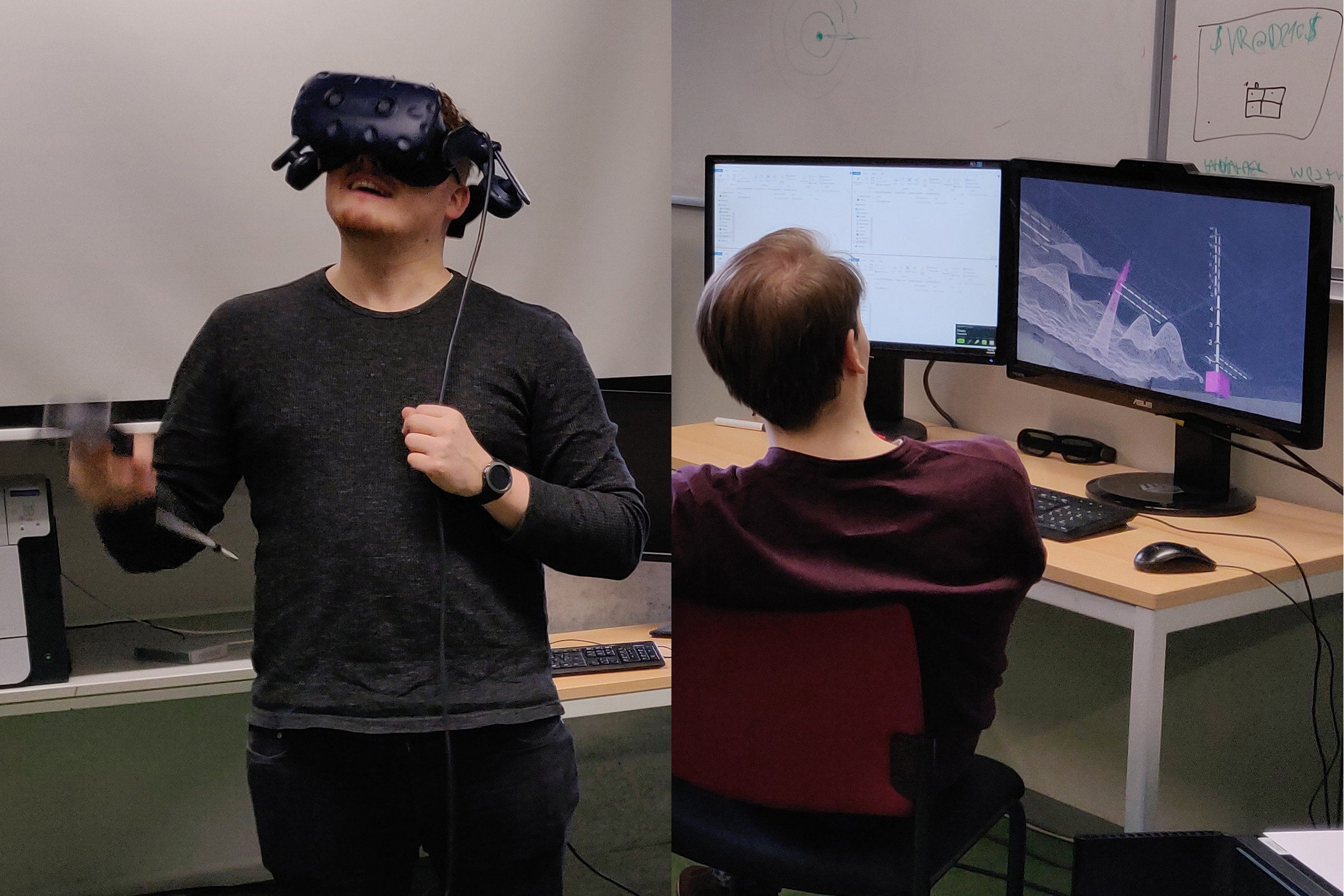} \\
    \small{Example posts at VisGuides (visguides.org)} &
    \small{A cluster of posts related to an assignment} &
    \small{An experiment on VR/VE inspired by VisGuides}
  \end{tabular}
  \caption{VisGuides is an online discussion forum dedicated to visualization guidelines \cite{Diehl:2018:EuroVis}. It has generated an interesting data set featuring discourse posts on various topics (a), and a cluster of posts related to an assignment in a visualization course (b). It has inspired us to categorize the data, apply text analytics to the data (b), and conduct empirical studies (c) according to grounded theory.}
	\label{fig:teaser}
}

\vgtcinsertpkg

\begin{document}


\firstsection{Introduction\label{sec:Introduction}}
\maketitle
%
\input{sections/intro.tex}
\input{sections/relwork.tex}
\input{sections/overview.tex}
\input{sections/coding.tex}
\input{sections/textanalytics.tex}
\input{sections/Swansea.tex}
\input{sections/KCL-interaction.tex}


\input{sections/Konstanz-VR.tex}


\section{Discussions and Conclusions}
\label{sec:Conclusions}
In this paper, we have reported several GT processes for grounding our understanding of several aspects of visualization guidelines on the data collected from VisGuides. Guided by GT, we developed a new categorization scheme for coding the data ($\S_{\ref{sec:VisGuides}}, \S_{\text{A}}, \S_{\text{B}}$), applied text analytics to the data ($\S_{\ref{sec:TextAnalytics}}, \S_{\text{C}}$), examined a subset of the data in an educational context ($\S_{\ref{sec:Teaching}}$), and conducted two empirical studies for in-depth theoretical sampling and comparative analysis ($\S_{\ref{sec:EmpiricalStudies}}$). These enabled us to obtain a collection of new findings, e.g., about:
\begin{itemize}
  \vspace{-2mm}
  \item how guidelines were discussed ($\S_{\ref{sec:VisGuides}}$),
  \vspace{-2mm}
  \item which guidelines were mentioned ($\S_{\ref{sec:VisGuides}}, \S_{\text{B}}$), 
  \vspace{-2mm}
  \item if the current topic categorization is suitable ($\S_{\ref{sec:VisGuides}}, \S_{\ref{sec:TextAnalytics}}, \S_C$),
  \vspace{-2mm}
  \item how guidelines were used in an educational context ($\S_{\ref{sec:Teaching}}$),
  \vspace{-2mm}
  \item the empirical evidence to support two specific guidelines on the topics of interaction and VR/VE respectively ($\S_{\ref{sec:EmpiricalStudies}}$).
\end{itemize}

We appreciated not only the systematic approach advocated by GT for grounding the answers to research questions on data collection, categorization, and analysis, but also the emphasis by GT for continuing investigation until theoretical saturation. In addition to further iterations for improving our categorization scheme when more data becomes available, further examination of the use of guidelines in different contexts, and further empirical studies for evaluating guidelines, we consider it desirable for the visualization community to make a collective effort to increase the mention and discussion of guidelines, especially in offering advice. One necessary step for enabling such effort is to address the challenge of creating a repository for storing guidelines and creating a suitable taxonomy or ontology for aiding the categorization and retrieval of guidelines.     


\bibliographystyle{abbrv-doi}

\bibliography{visguides-2019}


\newpage
~\\
\newpage

\appendix

\begin{center}
\LARGE
\textsf{\textbf{APPENDICES}}
\end{center}

\noindent\Large
\textsf{\textbf{Studying Visualization Guidelines\\
According to Grounded Theory}}\\

\noindent\normalsize
\textsf{Alexandra Diehl, \textit{Member, IEEE}, University of Konstanz}\\
\textsf{Matthias Kraus, \textit{Member, IEEE}, University of Konstanz}\\
\textsf{Alfie Abdul-Rahman, \textit{Member, IEEE}, King's College London}\\
\textsf{Mennatallah El-Assady, \textit{Member, IEEE}, University of Konstanz}\\
\textsf{Benjamin Bach, Edinburgh University}\\
\textsf{Robert S. Laramee, \textit{Member, IEEE}, Swansea University}\\
\textsf{Daniel A. Keim, \textit{Member, IEEE}, University of Konstanz}\\
\textsf{Min Chen, \textit{Member, IEEE}, University of Oxford}

\vspace{4mm}

\section*{\textbf{Open Data for Reproducibility}}
We plan to make all data used in the work available in public repositories in accordance with with the GDPR (The EU General Data Protection Regulation). 
{ We have already made the experiment data for Section 7.1 (the interaction study) and Section 7.2 (the VR/VE study) available at Github (\url{https://github.com/alfieabdulrahman/VisGuides_VIS2019} and \url{https://github.com/MatthiasKraus/Office-vs-Plain}). 
We are preparing the data for Section 4 (coded data) for releasing soon. 
All posts at VisGuides can be viewed by the public, so they are essentially open data. For releasing the XML version of the data, we will need to remove all personal data according to the GDPR. We are currently consulting GDPR experts as to what information can be included in the XML version of the data, and what has to be removed.

\acknowledgments{
We would like to thank all of those who have contributed to the discourses at the VisGuides Forum, including many visualization scientists and practitioners who have contributed their valuable time and expertise. We also thank Mohammad Alharbi (Swansea University) and Gemza Ademaj (University of Konstanz) for their assistance in the GT coding process.
}
\\

\input{sections/appendix-A.tex}

\input{sections/appendix-B.tex}

\input{sections/appendix-C.tex}


\input{sections/appendix-D.tex}

\input{sections/appendix-E.tex}



\end{document}

%% file: sections/intro.tex
A \emph{visualization guideline} embodies a wisdom advising a sound practice for designing and producing visualization imagery in practical applications.
In the field of visualization, it is estimated that hundreds of different guidelines have been recommended by various books, research papers, and online media \cite{kandogan2016grounded}.
These guidelines, if defined properly, are invaluable to both practical applications and the theoretical foundation of visualization.
As articulated in \cite{Chen:2017:CGA}, visualization guidelines can provide the taxonomies and ontologies in visualization with keywords and causal relations.
Each guideline can be mapped to a mathematical conjecture, and it can potentially become a theorem when it is mathematically confirmed.
A collection of related guidelines (e.g., on the topic of using colors in visualization) can inform the development of a conceptual model (e.g., about color perception in visualization).

Most visualization guidelines were proposed based on authors' experience in designing and producing visualization imagery, or on findings obtained from various forms of empirical studies.
It is inevitable that some of the guidelines may not be applicable in all contexts or conditions, and some may be in conflict with one another.
It is not a trivial undertaking to define every guideline precisely, together with specifications of those applicable conditions as well as those inapplicable.
Nevertheless, if visualization is to play a credible role in any mission-critical data analytics, knowledge discovery, and decision making workflow, guidelines for producing visualizations must be rigorously evaluated, critiqued, and maintained, for example, in the same way that medical or healthcare guidelines would be scrutinized.

A guideline can be considered as a postulated theory emerged from practical experience or empirical studies. In social science, \emph{grounded theory} (GT) is a methodology to ``scrutinize'' a postulated theory.
The methodology, which was first proposed by Glaser and Strauss for sociological research \cite{Glaser:1967:book}, asserts that a postulated theory should be ``grounded'' in the data space where the theory emerges, and stipulates a collection of methods and processes, such as \emph{categorization}, \emph{coding}, \emph{constant comparative analysis}, \emph{negative case analysis}, \emph{memoing}, and so on. The ultimate goal is to scrutinize the theory until the effort has reached \emph{theoretical saturation}.
Clearly, this methodology is, in principle, applicable to visualization guidelines.

In this paper, we present a number of research activities for studying visualization guidelines, which were organized around and benefited from VisGuides (visguides.org), which is a web-based forum for discussing visualization guidelines \cite{Diehl:2018:EuroVis}.
It naturally provides a valuable source of data where guidelines (as postulated theories) can be grounded.
In this work, we conducted several research activities that represent different methods and processes stipulated by GT, aiming to ground the following six research questions in the data:
\begin{description}
\vspace{-2mm}
\item [Q1.] Whether and in which way were visualization guidelines mentioned in the posts at VisGuides?
\vspace{-2mm}
\item [Q2.] Which visualization guidelines were mentioned and how did the relevant posts discuss them?
\vspace{-2mm}
\item [Q3.] Does the current topic categorization for visualization guidelines at VisGuides work satisfactorily?
\vspace{-2mm}
\item [Q4.] Did VisGuides supplement and enhance the learning experience of those studying the topic of data visualization?
\vspace{-2mm}
\item [Q5.] What visualization guidelines mentioned could be included in the lectures on data visualization?
\vspace{-2mm}
\item [Q6.] What is the evidence for two specific guidelines on interaction and VR/VE (virtual reality/environments) respectively? 
\end{description}
%
%
With this multifaceted effort to study visualization guidelines according to GT, we have obtained a number of findings about the aforementioned questions, which are to be detailed in Section \ref{sec:VisGuides} on categorzation and coding, Section \ref{sec:TextAnalytics} on text analytics, Section \ref{sec:Teaching} on visualization education, and Section \ref{sec:EmpiricalStudies} on two empirical studies.
Our contributions include:
\begin{itemize}
  \vspace{-1.5mm}
  \item We have made a major multifaceted effort to study visualization guidelines according to GT, representing a significant step towards an organized approach to studying visualization guidelines as part of the theoretical foundation of visualization \cite{Chen:2017:CGA}.
  \vspace{-1.5mm}%
  \item We have shown that we can enrich and enhance the methods and processes instigated by GT by using digital technologies (e.g., a web-based forum, data visualization, and text analysis) as well as empirical studies and educational activities, all of which are rarely reported in traditional research work on GT.
  \vspace{-1.5mm}%
  \item Using the data obtained in this work, we have provided further evidence to support two visualization guidelines, uncovered a number of guidelines buried in the literature, and identified several challenges and opportunities for attaining \emph{theoretical saturation} in studying visualization guidelines. 
\end{itemize}

%% file: sections/relwork.tex
\section{Related Work}
\label{sec:RelatedWork}

A guideline is a ``general rule, principle, or piece of advice'' \cite{guidelineoxford} and provides ``information intended to advise people on how something should be done or what something should be'' \cite{guidelinecambridge}.
Kandogan and Lee reported that they identified ``550 [visualization] guidelines, ranging in length from 5 to 140 words'' \cite{kandogan2016grounded}.
Several surveys collected a number of guidelines, including uses of colors in information visualization \cite{silva2011using}; design guidelines for interactive visualizations \cite{brodbeck2009interactive}, glyph-based visualizations \cite{borgo2013glyph}, and visualization in scientific publications \cite{kelleher2011ten}; and design patterns for storytelling \cite{bach2018narrative,bach2018design}.
 
A good number of these guidelines are known to visualization researchers and practitioners, such as ``the rainbow color map is harmful'' \cite{Rogowitz:1998:S,Borland:2007} and ``maximize data-ink ratio'' \cite{tufte2001visual}.
Many guidelines, which may not be explicitly phrased as guidelines, can be extracted from research findings and theoretical propositions. For example Bertin's findings on the suitability of different retinal variables (visual channels) for four types of perception \cite{Bertin:1983:book} can easily be rephrased as ``do'' and ``don't'' statements.     
Some of these guidelines have been widely endorsed, while only a few have been mathematically confirmed. Shneiderman's ``overview first, zoom and filter, then details-on-demand'' \cite{Shneiderman:1996:VL} has been widely adopted, and the ``zoom'' part has been mathematically proved \cite{Chen:2010:TVCG}. It was the discoveries of the occasional preference in some applications for ``details first'' \cite{vanHam:2009:TVCG,Luciani:2019:TVCG} that led to better understanding of its underlying rationale \cite{Chen:2016:book}. This indicates the importance to ground visualization guidelines on practical experience.

Meyer et al. proposed to use visualization guidelines to help practitioners make choices
in developing visual designs and visualization systems \cite{Meyer:2015:IV}. Zuk et al. proposed to use guidelines as heuristics for evaluation \cite{Zuk:2006:BELIV}.
Engelke et al. proposed a supply chain framework for the life-cycle of guidelines  
\cite{engelke2018visupply}.
Scott-Brown discussed ways for presenting guidelines \cite{Scott-Brown2018Presenting}.
Chen et al. asserted that guidelines are an integral part of the theoretical foundation of visualization and instigated the need for curating, evaluating,
critiquing, and refining guidelines in an open and transparent manner (including adopting GT) \cite{Chen:2017:CGA}. Bradley et al. suggested that visualization research can benefit from a range of research methods in humanities and social science (including GT) \cite{Bradley:2018:CGA}.
Sprague and Tory proposed to build a grounded evaluation methodology upon GT  \cite{Sprague:2012:IV}.
Chandrasegaran et al. used visual analytics to aid ground theory processes \cite{Chandrasegaran:2017:CGF}.
This work follows the research agenda on visualization guidelines envisaged by these works.

Kandogan and Lee conducted a GT study on the language characteristics of visualization guides \cite{kandogan2016grounded}. They have identified five high-level concepts (i.e., Data, Visualization, User, Insight, and Device) and a refining/contextualizing concept (Qualifier) that a guideline may feature. They have also identified eight 1st-level categories and eight 2nd-level categories for characterize the language features used by guidelines for conveying different relationships. In addition, they also considered language forms for conveying instructions, conditions, rankings, and explanations. As the data used in \cite{kandogan2016grounded} is not publicly available, it is not easy to build further ground theory processes directly on such a substantial collection of visualization guidelines.
This work took a different research direction by focusing on the open discourses on guidelines and their uses in practical scenarios. Instead of using books and papers as data sources, we harvested data from the online discussion forum VisGuides \cite{Diehl:2018:EuroVis}. The categorization scheme derived from our GT processes thus characterize a set of features that are not typical in books and papers. In addition, we also used text analysis and visualization complement the traditional GT processes and conducted empirical studies to ground two selected visualization guidelines further into data.   

%% file: sections/overview.tex
\begin{table*}[t!]
  \centering
  \caption{A summary of the principles and methods of grounded theory, and an outlook on how computer-assisted implementation may complement the traditional implementation of the grounded theory (GT) methodology \cite{Willig:2013:book}.
  The section tags $\S_{\ref{sec:VisGuides}}$, $\S_{\ref{sec:TextAnalytics}}$, $\S_{\ref{sec:Teaching}}$, and  $\S_{\ref{sec:EmpiricalStudies}}$ indicate those processes that were used in this work and detailed in the corresponding sections. The processes in \emph{italic} indicate possible future work. See also Appendix D.}
  \label{tab:GroundedTheory}
  
  \begin{tabular}{@{}p{3.2cm}@{\hspace{4mm}}p{7cm}@{\hspace{4mm}}p{7.2cm}@{}}
  \textbf{Methods and Principles} & \textbf{Traditional Implementation} &  \textbf{Computer-assisted Implementation}\\\toprule
  \textbf{Categorization}
  & Close reading ($\S_{\ref{sec:VisGuides}}$, $\S_{\ref{sec:Teaching}}$),
    category identification ($\S_{\ref{sec:VisGuides}}$, $\S_{\ref{sec:TextAnalytics}}$, $\S_{\ref{sec:Teaching}}$);
    \emph{taxonomy construction}; \emph{ontology construction}.
  & Distant reading with statistics and statistical graphics ($\S_{\ref{sec:VisGuides}}$);
    cluster and similarity analysis ($\S_{\ref{sec:TextAnalytics}}$);
    text visualization ($\S_{\ref{sec:TextAnalytics}}$);
    \emph{topic modelling};
    \emph{algorithmic taxonomy construction};
    \emph{algorithmic ontology construction}. \\\midrule
  \textbf{Coding}
  & Close reading and labelling ($\S_{\ref{sec:VisGuides}}$, $\S_{\ref{sec:Teaching}}$); open coding ($\S_{\ref{sec:VisGuides}}$); axial coding ($\S_{\ref{sec:VisGuides}}$); selective coding ($\S_{\ref{sec:VisGuides}}$, $\S_{\ref{sec:Teaching}}$).
  & Text and discourse analysis ($\S_{\ref{sec:TextAnalytics}}$);
    text visualization ($\S_{\ref{sec:TextAnalytics}}$);
    similarity analysis ($\S_{\ref{sec:TextAnalytics}}$);
    network visualization ($\S_{\ref{sec:TextAnalytics}}$);
    \emph{association analysis}; \emph{ontology mapping}. \\\midrule
  \textbf{Comparative Analysis}
  & Iterative close reading ($\S_{\ref{sec:VisGuides}}$);
    comparing different options of category abstraction ($\S_{\ref{sec:VisGuides}}$);
    continuous refinement of the categorization scheme ($\S_{\ref{sec:VisGuides}}$);
    proposing new categories or categorization schemes ($\S_{\ref{sec:VisGuides}}$). 
  & Iterative and continuous effort for distant reading with statistics and statistical graphics; cluster and similarity analysis; topic modelling; text visualization; algorithmic taxonomy and ontology construction.\\\midrule
  \textbf{Negative Case Analysis}
  & Close reading ($\S_{\ref{sec:VisGuides}}$, $\S_{\ref{sec:Teaching}}$);
    negative case identification and analysis ($\S_{\ref{sec:VisGuides}}$, $\S_{\ref{sec:Teaching}}$). 
  & Algorithmic outlier analysis and anomaly detection.\\\midrule
  \textbf{Memoing}
  & Memo-writing ($\S_{\ref{sec:VisGuides}}$, $\S_{\ref{sec:Teaching}}$);
    sketch-drawing ($\S_{\ref{sec:VisGuides}}$, $\S_{\ref{sec:Teaching}}$);
    \emph{systematically recording the ideas and actions related to the development of a theory}. 
  & Online discussion forum ($\S_{\ref{sec:VisGuides}}$, $\S_{\ref{sec:Teaching}}$);
    crowd sourcing ($\S_{\ref{sec:VisGuides}}$, $\S_{\ref{sec:Teaching}}$);
    provenance visualization ($\S_{\ref{sec:Teaching}}$);
    \emph{data collection from social media}. \\\midrule
  \textbf{Theoretical Sensitivity}
  & Testing a theory against old and new data (i.e., ``Interact'' with data) ($\S_{\ref{sec:VisGuides}}$, $\S_{\ref{sec:Teaching}}$, $\S_{\ref{sec:EmpiricalStudies}}$);
    analysis of applicability, negative cases, and possible new categories ($\S_{\ref{sec:VisGuides}}$). 
  & Computer-assisted data collection ($\S_{\ref{sec:VisGuides}}$, $\S_{\ref{sec:Teaching}}$, $\S_{\ref{sec:EmpiricalStudies}}$),
    data analysis and visualization ($\S_{\ref{sec:VisGuides}}$, $\S_{\ref{sec:TextAnalytics}}$, $\S_{\ref{sec:Teaching}}$, $\S_{\ref{sec:EmpiricalStudies}}$). \\\midrule
  \textbf{Theoretical Sampling}
  & Data collection ($\S_{\ref{sec:VisGuides}}$, $\S_{\ref{sec:Teaching}}$, $\S_{\ref{sec:EmpiricalStudies}}$); empirical studies (e.g., controlled experiments ($\S_{\ref{sec:EmpiricalStudies}}$), \emph{surveys, focus groups, field observation, etc.});
  \emph{interview transcription}.
  & Online discussion forum ($\S_{\ref{sec:VisGuides}}$, $\S_{\ref{sec:Teaching}}$),
    crowd sourcing ($\S_{\ref{sec:VisGuides}}$, $\S_{\ref{sec:Teaching}}$), and
    \emph{data collection from social media};
    computer-assisted experiments ($\S_{\ref{sec:EmpiricalStudies}}$);
    computer-mediated group activities ($\S_{\ref{sec:Teaching}}$);
    observation of online activities ($\S_{\ref{sec:VisGuides}}$, $\S_{\ref{sec:Teaching}}$). \\\midrule
  \textbf{Theoretical Saturation}
  & \emph{Data collection and analysis until theoretical saturation has been achieved}.
  & Computer-assisted platforms for facilitating the longevity and provenance of the GT processes ($\S_{\ref{sec:VisGuides}}$). \\\bottomrule
  \end{tabular}
  \vspace{-4mm}
\end{table*}

\section{Overview and Methodology}
\label{sec:GroundedTheory}
\emph{Grounded theory} (GT) was proposed in the context of the social sciences \cite{Glaser:1967:book}. Table \ref{tab:GroundedTheory} lists its main methods (\emph{categorization}, \emph{coding}, \emph{comparative analysis}, \emph{negative case analysis}, and \emph{memoing}), and principles (\emph{theoretical sensitivity}, \emph{sampling}, and \emph{saturation}). The formal descriptions of these methods and principles, which was compiled based on \cite{Willig:2013:book}, can be found in Appendix D. 

Consider a postulated theory or a research question (e.g., about social behaviors), which may be in the form of a categorization scheme, a taxonomy, an ontology, a causal relation, a conceptual model, or a numerical model. GT instigates that the theory needs to be grounded in the data space where the theory emerged, by constantly and continuously sampling the data space, analyzing the data captured, and refining the theory until it reaches theoretical saturation. Here we differentiate a \emph{data space} from a \emph{data set}. The former consists of all possible data sets in a context (e.g., all possible posts that could appear at VisGuides), while the latter is a set of data points as an instance of sampling the data space (e.g., the actual posts in a particular period).

\vspace{-1mm}
\paragraph{\textbf{Grounding Visualization Guidelines on Data.}}
From the perspective of GT, visualization guidelines are theories that should be grounded on data, since almost all these guidelines were proposed based on the originators' observation and experience, or on the results of empirical studies.
GT instigates to orient research questions towards what happened (e.g., ``How do people do X?'') rather than mental states (e.g., ``What do people want?'' or ``Why do people do X?'' \cite{Willig:2013:book,Strauss:1998:book}.
We therefore formulated all our six research questions in this work (as described in Section \ref{sec:Introduction}) as ``What happened?'' questions.

In addition to considering visualization guidelines as individual theories, we can anticipate the possibility of formulating meta-theories that model different aspects of the corpus of visualization guidelines, e.g., their categorization \cite{kandogan2016grounded} and their lifecycle \cite{engelke2018visupply}.
According to GT, such meta-theories should also be grounded on the relevant data, which may include different expressions of the guidelines, the attributes of each expression (e.g., originator, location, period, context, etc.), and documentation about their formulation, discussion, curation, sharing, application, evaluation, refinement, and so on. 

\vspace{-1mm}
\paragraph{\textbf{VisGuides Data.}}
VisGuides (\url{visguides.org}) is an online discussion forum dedicated to visualization guidelines \cite{Diehl:2018:EuroVis}.
It allows registered users to propose and recommend guidelines, pose questions and offer advice, share positive and negative experience about certain guidelines, report and reason about successes, failures, and conflicts of guidelines, and suggest ways to refine guidelines.
VisGuides was launched on September 27, 2017 \cite{Diehl:2018:EuroVis}. There were 248 posts up to March 13, 2019 when we collected the latest data set $D_{20190313}$ for the analysis in this paper. The data set contains discourses on some prevalent topics (e.g., color, interaction, virtual reality (VR), and so on), and contributions from users with many different visualization backgrounds, ranging from undergraduate students to highly-respected visualization experts (e.g., in alphabetic order, Niklas Elmqvist, Stephen Few, Eduard Gr\"{o}ller, Theresa-Marie Rhyne, Bernice Rogowitz, and so on).

It is a digital platform that features both the ``theories'' (e.g., visualization guidelines) to be developed, and the ``data'' (i.e., discourses) that can be used to generate, evaluate, and refine the theories.
Following the tradition in the field of visualization, we use the terms such as ``research questions'', ``hypotheses'', ``categorization schemes'', and ``findings'' instead of ``theories'' in the following discussions.

In the following four sections, we describe four strands of research activities inspired and guided by GT. While we employed traditional GT processes, we also utilized our own expertise in computer-assisted data analysis and visualization to complement the traditional processes. These activities are summarized below, where the same tags used in Table \ref{tab:GroundedTheory}, such as $\S_{\ref{sec:VisGuides}} \backsim \S_{\ref{sec:EmpiricalStudies}}$ and $\S_{\text{A}} \backsim \S_{\text{E}}$, relate the GT processes to individual sections and appendices.

\begin{itemize}
    \vspace{-2mm}%
    \item \textbf{Traditional Categorization and Coding ($\S_{\ref{sec:VisGuides}}$, $\S_{\ref{sec:Teaching}}$, $\S_{\text{A}}$, $\S_{\text{B}}$).}
    The data set $D_{20190313}$ collected from VisGuides consists of primarily free text, any answers to the six research questions \textbf{Q1-Q6} depend critically on the interpretation of the text. According to GT, the key to befitting interpretations is a correct \textbf{categorization} and \textbf{coding}.
    We close-read all posts in the data set $D_{20190313}$, carried out open and axial coding in order to derive categorization schemes as candidate theories. We used selective coding and statistical graphics to evaluate these categorization schemes. We then proposed an integrated categorization scheme that allowed us to obtain several findings.
    \vspace{-2mm}%
    \item \textbf{Text Analysis and Visualization (Section {\ref{sec:TextAnalytics}}, $\S_{\ref{sec:TextAnalytics}}$, $\S_{\text{C}}$).}
    In conjunction with the traditional approach, we applied several text analysis and text visualization techniques to $D_{20190313}$, which allowed us to observe many patterns that could not be derived from close reading easily.
    \vspace{-2mm}%
    \item \textbf{Visualization Coursework (Section \ref{sec:Teaching}, $\S_{\ref{sec:Teaching}}$).} With hundreds of guidelines available in the literature \cite{kandogan2016grounded}, it would take a very long time for an individual guideline to receive a good number of hits in sampling. When VisGuides was used to support a piece coursework, we took the opportunity to define a specific context in which the number of applicable guidelines is much smaller and the sampling can be much more intensive. We denote this subset of data as $D_{\text{Swansea2019}} \subset D_{20190313}$. The close-reading and coding of this $D_{\text{Swansea2019}}$ revealed further findings.
    \vspace{-2mm}%
    \item \textbf{Empirical Studies (Section \ref{sec:EmpiricalStudies}, $\S_{\ref{sec:EmpiricalStudies}}$).}
    When our initial examination of two specific guidelines indicated a lack of data, we followed the GT methodology to collect more data outside VisGuides. We conducted two empirical studies that resulted in two additional data sets $E_{\text{KCL2019}}$ and $E_{\text{Konstanz2019}}$. These data sets provided the two guidelines with supporting evidence.      
\end{itemize}

%% file: sections/coding.tex
\section{Categorization and Coding}
\label{sec:VisGuides}
\newcommand\perc[1]{\the\numexpr#1*100/217\%}
\newcommand{\says}[1]{\textit{`#1'}}


In this section, we focus on several holistic research questions on visualization guidelines, including \textbf{Q1-Q3} stated in Section \ref{sec:Introduction}.

Because VisGuides is a computer-assisted platform, the computer has to some extent provided some categorization schemes and automated coding (or selective coding in a grounded theory term). These include those variables in Table \ref{tab:Computer} in Appendix A for coding each post. Clearly, these categorization variables would not be sufficient for supporting our investigation into the aforementioned research questions. 

\newcommand{\xx}{\textcolor{red}{XX}}
\vspace{-1mm}
\paragraph{\textbf{The GT Coding Process.}}
We used the GT methodology to formulate a categorization scheme for the data set $D_{20190313}$, and used the scheme to code all 248 posts in the data set. Two researchers acted as the main coders with the help of six other researchers. It took three iterations of \emph{open coding}, \emph{axial coding}, and \emph{selective coding} for us to derive the main categorization scheme with six variables (version 3), which was revised slightly to yield version 4.
The details of our GT coding process can be found in Appendix A.4.

\begin{figure}[t]
    \centering
    \includegraphics[width=1\columnwidth]{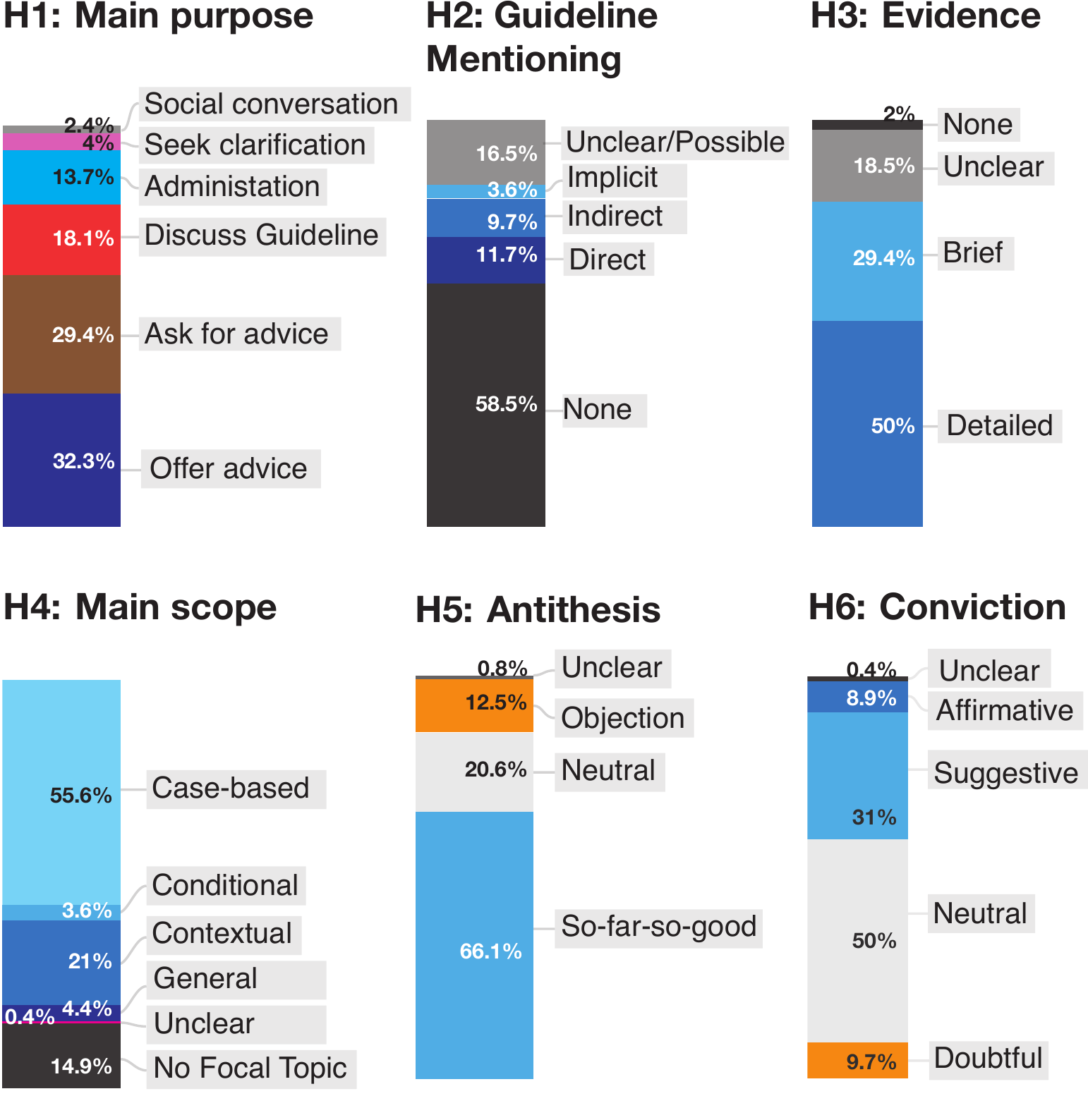}
    \caption{Distribution of the categories of the 248 posts in $D_{20190313}$.}
    \label{fig:catstats}
    \vspace{-4mm}
\end{figure}

\paragraph{\textbf{Variables and Categories.}}
Our GT coding process resulted in a categorization scheme that is described in detail in Appendix A. For the self-containment, we briefly summarize the scheme here. The scheme consists of six variables, which include:
\begin{description}[noitemsep]
  \item [H1.] \textbf{Main Purpose}: This variable captures the main purpose of a post. If the post exhibits more than one purpose, the coder made a judgment about what the main one is. Its categories include \emph{Ask for Advice}, \emph{Offer Advice}, \emph{Discuss Guidelines}, \emph{Seek Clarification} (e.g., ``Please explain X'', \emph{Add Classification}, \emph{Social Conversation} (e.g., ``Thank you''), and \emph{Administration}.
  \item [H2.] \textbf{Guideline Mentioning}: This describes if and how a post mentions a visualization guideline. Its categories include \emph{Direct} (a guideline is explicitly mentioned in the post), \emph{Indirect} (a reference is given and it likely contains one or more guidelines related to the post), \emph{Implicit} (a reasonably well-known guideline can be identified by the coder), \emph{Unclear or Possible} (the coder is uncertain), and \emph{None} (the coder is certain that there is no guideline).
  \item [H3.] \textbf{Evidence}: This grades the level of detail of the supporting evidence in a post. Its categories include \emph{Detailed}, \emph{Brief} (typically relying on references), \emph{Unclear}, and \emph{None}.
  \item [H4.] \textbf{Main Scope}: This characterizes the generality of the discourse in the post in terms of its applicability. Its categories include \emph{Case-based} (focusing on one or more individual cases), \emph{Conditional} (applicable to a number of cases under certain conditions), \emph{Contextual} (focusing on a high-level context, e.g., a topic, a type of tasks, a type of users, etc.), \emph{General} (applicable to most visualization contexts), \emph{Unclear}, and \emph{None}.
  \item [H5.] \textbf{Antithesis}: This reflects the stance of a post with respect to all previous posts in the thread. Its categories include \emph{Neutral} (no opinion expressed), \emph{So Far So Good} (agreeing with all previous posts), \emph{Objection} (disagreeing with at least one previous posts), and \emph{Unclear}.
  \item [H6.] \textbf{Conviction}: This classifies the level of conviction shown by the arguments in the post. Its categories include \emph{Neutral} (no opinion expressed), \emph{Doubtful} (expressing some doubts), \emph{Suggestive} (suggesting an opinion with some uncertainty), \emph{Affirmative}, and \emph{Unclear}.
\end{description}
\vspace{-1mm}
In addition to these six variables, we have extracted from all directly- and indirectly-mentioned guidelines from the 248 posts. They are listed in Appendix B. 

\begin{figure}[t]
    \centering
    \includegraphics[width=1\columnwidth]{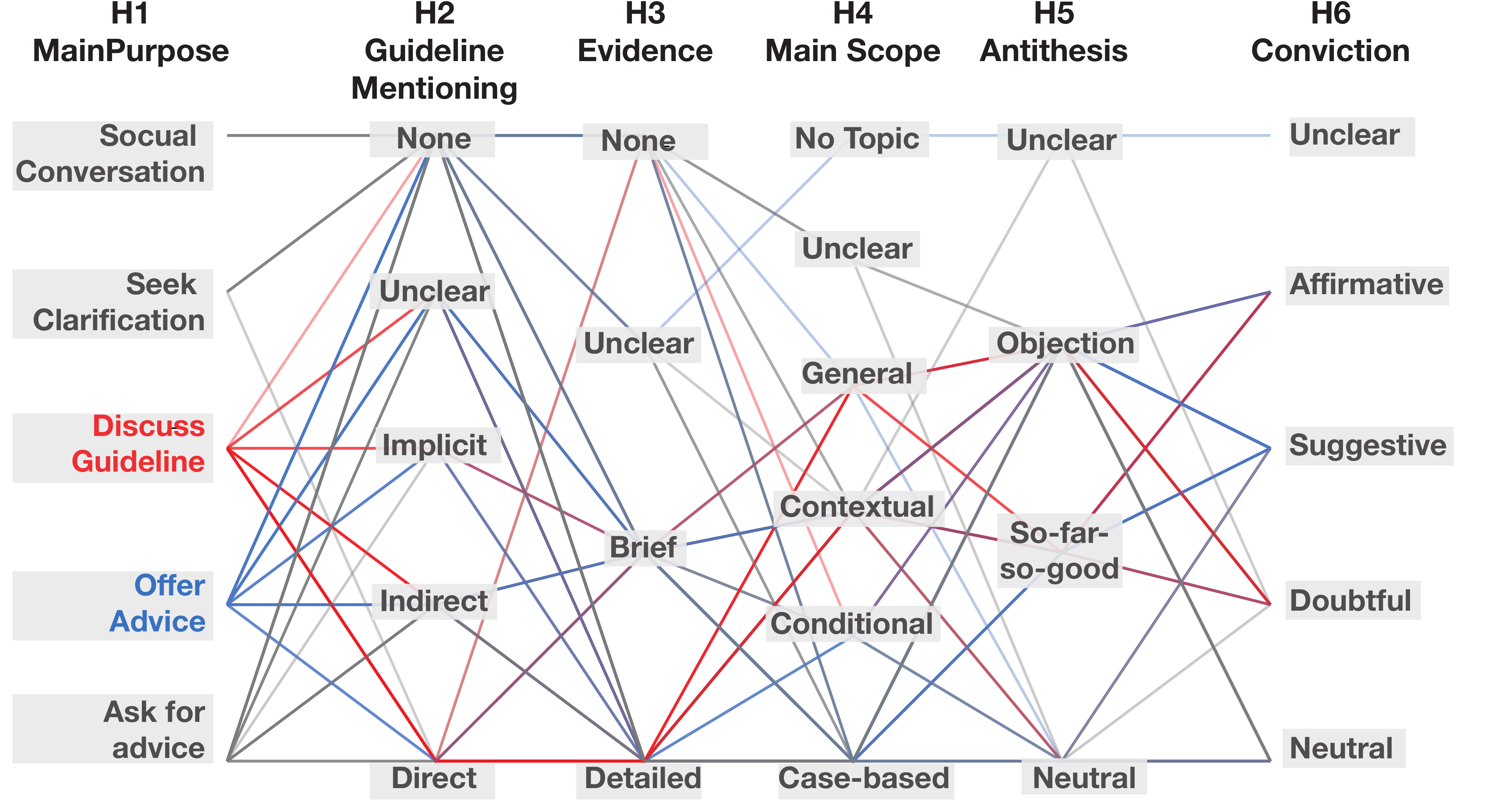}
    \caption{Visualizing the relationships among different variables. Red lines are associated with the category of \emph{Discuss Guidelines}, while the blue lines are associated with that of \emph{Offer Advice}.}
    \label{fig:catpcp}
    \vspace{-4mm}
\end{figure}

\vspace{-1mm}
\paragraph{\textbf{Results and Findings.}}
Fig. \ref{fig:catstats} shows the summary statistics of the 248 posts in the data set $D_{20190313}$ in terms of their categorization based on the six variables, while Fig. \ref{fig:catpcp} depicts the relationships among different variables. From the coded data and the corresponding statistics and visualization, we have made the following observations.

In terms of \textbf{Q1} (Whether and in which way were visualization guidelines mentioned in the posts at VisGuides?), Fig. \ref{fig:catstats} shows that some 29.4\% of the posts are in the H1 categories of \emph{Ask for Advice} and \emph{Offer Advice}, while about 18.1\% are in the category of \emph{Discussing Guidelines}. About 58.5\% of the posts do not mention any guideline at all (H2), much higher than the 21.4\% that mentioned guidelines directly and indirectly.
Exactly half of the posts (50.0\%) focused on specific cases (H3), e.g., about a specific design. The discussions on guidelines typically are mostly contextual (H4, 21.0\%) rather than conditional (3.6\%). In other words, it has been rare to discuss the conditions where a guideline may or may not be applicable.
Fig. \ref{fig:catpcp} compares how guidelines are mentioned by posts that \emph{Offer Advice} (blue lines) and those that \emph{Discuss Guidelines} (red lines), revealing that most posts offering advice do not mention guidelines neither directly nor indirectly. 
Our finding indicates that \textbf{using and discussing guidelines is not as common as one would prefer to see.} We conjectured a few possible reasons: (i) recalling guidelines precisely may be difficult without a searchable repository; (ii) discussing guidelines may require broad or in-depth knowledge about visualization; (iii) applying some guidelines to practical problems may be easier said than done when one does not know the applicable conditions of a guideline.

Also in terms of \textbf{Q1}, our close-reading indicated that many posts that \emph{Offer Advice} contained sound insight and detailed explanations about the reasons behind an advice. This is very positive given that posting at VisGuides is voluntary. Many of such posts contained examples and explained the reasons behind the advice. Of course, in some cases, a brief answer is everything it takes to solve a problem. However, only a few posts contained references to the sources of guidelines.
The most common reference was to ColorBrewer \cite{brewer1994color}, which perhaps has a quasi-guideline status due to its wide acceptance and utilization. We also found that it is rare for posts to propose new guidelines, or make abstraction or generalization from individual scenarios. 
Out finding indicates that \textbf{it is easy to talk about details, taxing to mention references, and difficult to abstract and generalize.} Our conjectures include the aforementioned reasons (i) and (ii).

Meanwhile, in terms of \textbf{Q2} (Which visualization guidelines were mentioned and how did the
relevant posts discuss them?), Appendix B shows that a total of 14 guidelines were directly mentioned, and another 15 references were mentioned as support to some discussions.
Fig. \ref{fig:catstats}(H5) shows that some 66.1\% of the posts did not voice any objection against any opinions in the previous posts, while only 12.5\% featured disagreement (objection). About half of the posts were shown to be neutral in offering opinions, while those that did offer views typically used a suggestive tone (H6, 31.1\%).
Our finding indicates that \textbf{posing a critical question or offering a critical comment is not common.} Our conjectures include the aforementioned reason (ii), and (iv) culturally many may be reluctant to challenge the status quo.

In terms of \textbf{Q3} (Does the current topic categorization for visualization guidelines
at VisGuides work satisfactorily?), the statistics derived from computer-coded topic categories shows that most posts are on the topic of ``Visual Design'' (79.8\%). The rest of the posts are sparsely distributed among the other 10 topic categories, including ``Perception'' (6.0\%), ``Education'' (2.8\%), ``Theory'' (2.4\%), ``Interaction'' (2.0\%), ``Uncategorized'' (2.0\%), ``Cognition'' (1.6\%), ``General'' (1.2\%), ``VR/VE'' (0.8\%), ``Site Feedback'' (0.8\%), and ``Medical Visualization'' (0.2\%). VisGuides offered a few subcategories under ``Visual Design'', each attracted 2.0\% or less.
This clearly indicates that \textbf{the current topic categorization is not adequate.} To our best knowledge, there has not been a topic categorization scheme proposed for visualization guidelines in the literature. Further research to develop taxonomies or ontologies of visualization guidelines is urgently needed.

%% file: sections/textanalytics.tex
\section{Text Analysis and Visualization}
\label{sec:TextAnalytics}

Considering that the data set $D_{20190313}$ is a text data set collected in a visualization context, one cannot resist the idea of applying text analysis and visualization to the data set. While the GT process for formulating categorization schemes and manually coding each post through close reading is scientifically rewarding and practically unavoidable, text analysis and visualization may potentially reveal patterns that close reading cannot discover. In many ways, we can consider text analysis and visualization is a form of ``open coding'' by computers. As long as human analysts are involved in comparative analysis and knowledge discovery, text analysis and visualization can complement the traditional GT methodology. 

\begin{figure}
    \centering
    \includegraphics[width=\columnwidth]{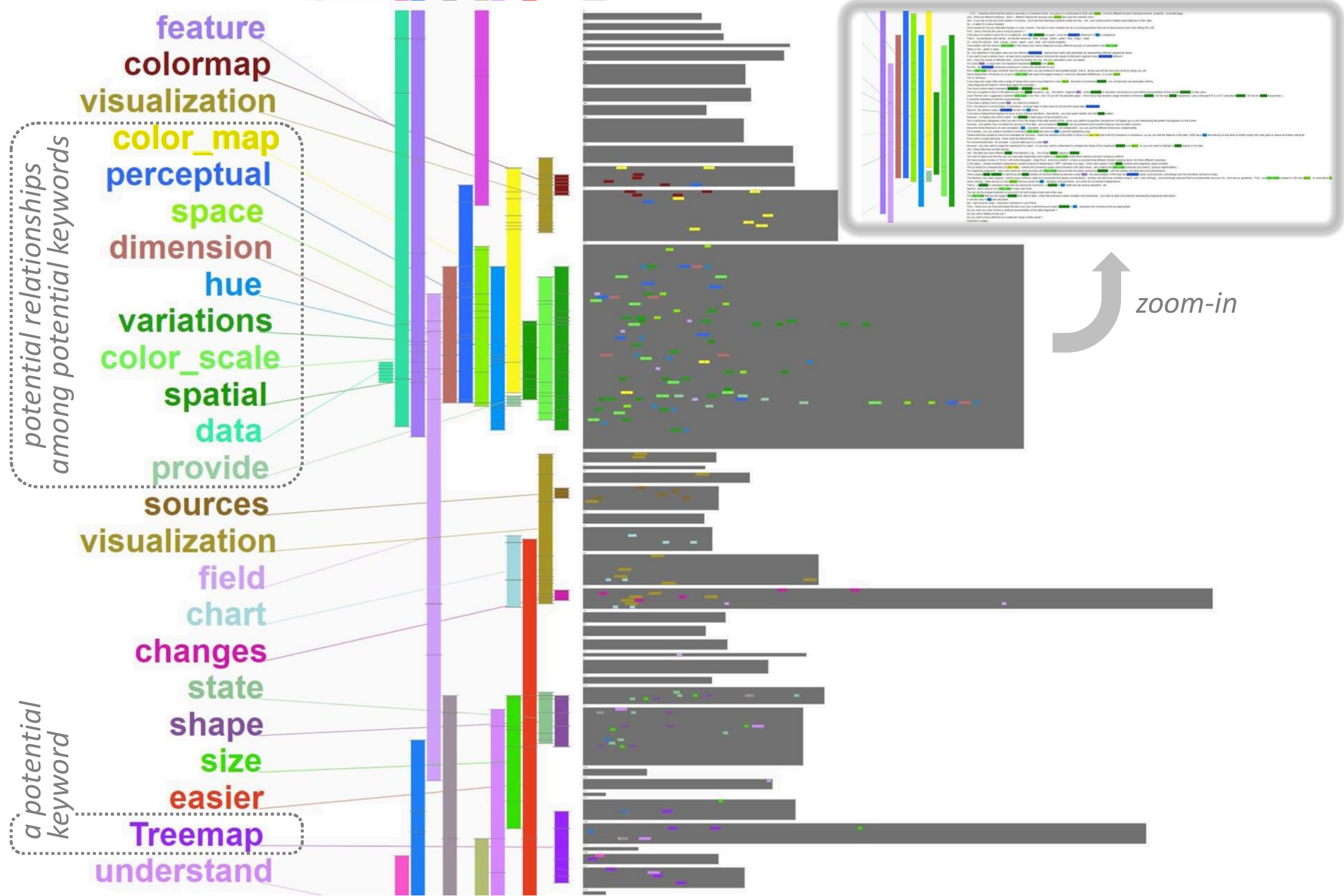}
    \caption{ The Lexical Episode Plot~\cite{Gold15} depicts the evolution of the discussion over time. It automatically detects \textit{compact chains of $n$-grams} and highlights them beside the complete text as an \textit{overview}. We can interactively zoom-in to the interesting text areas for \textit{close-reading}.}
    \label{fig:LEP}
    \vspace{-4mm}
\end{figure}

The analysis of forum discussions is an active area of research~\cite{Liu2018} with the goals of capturing the dynamics of topics, participants, as well as the linguistic quality of the posts and utterances. Such a type of text data bares the characteristics of verbatim text, i.e., it is based on a social interactions, with a rapid creation cycle and evolution. In contrast to highly-edited text, commonly found in traditional print media, such as books and articles, this type of text brings along the challenge of being more noisy and heterogeneous~\cite{Hautli-Janisz2017a}. Hence, to analyze such corpora, a tailored set of text analytics tools is required. To tackle such a challenge, the \textit{VisArgue platform}~\cite{El-Assady2017c} has been developed as the first linguistic visualization framework specialized on analyzing conversational text, such as in political debates and forums. It integrates a multi-step linguistic pipeline with state-of-the-art statistical processing and machine learning to provide a wide set of visualizations as perspective on the analyzed data. For instance, for the analysis of forum data, the \textit{ThreadReconstructor}~\cite{El-Assady2018b} visualization offers a technique for untangling threaded discussions. On the other hand, \textit{sequential pattern mining}~\cite{DBLP:conf/vissym/JentnerEGK17} techniques are utilized to reveal patterns in parallel conversational corpora. 
Furthermore, to reveal the dynamics between conversation participants, \textit{ConToVi}~\cite{EGA+16a} applies topic modeling, as well as micro-linguistic feature detection, highlighting the argumentation and language use styles of speakers.  

From the data set $D_{20190313}$, we extracted 84 posts.
For each thread, XML structure of $D_{20190313}$ consists of an initial question and subsequent replies.
We removed the URLs, images, and stop-words from the text and eliminated the posts that contain only administrative messages and social conversations.
We then employed several techniques to reveal the \textit{contextual relationships} among these posts.
The main goal of our visual text analysis is to tackle the challenge of categorizing visualization guidelines. Based on the statistical analysis, we deduced that the 11 current categories of guidelines of VisGuides are not ideal.
We thus consider the data-driven text analysis to reveal a more informed categorization. We further use visual text analytics to study the dynamics between participants in the discourse based on sentiment relations in the underlying text collection.   

We first used \textit{Lexical Episode Plots}~\cite{Gold15} to analyze the evolution of the discussions. This technique extracts densely occurring lexical chains, as visualized in \autoref{fig:LEP}. Objects are placed on a zoomable canvas to enable a smooth transition between close- and distant-reading. Threads are placed in chronological order (from top to bottom) and abstracted with gray blocks for distant-reading. Each of the blocks dissolves into its respective text when zoomed-in to enable close-reading. On the left-hand side, the extracted \textit{episodes} are placed. These are $n$-grams that occur more densely than expected if assuming a uniform distribution of words. Every occurrence of an \textit{episode} is marked by a horizontal line on its respective bar, as well as a colored bar at the word's position in the thread.
From \autoref{fig:LEP}, a human coder can observe a number of keywords (e.g., \textit{colormap}, \textit{spatial}, and \textit{treemap}) that may be considered potentially as topics. 
It also shows a potential co-occurrence relationship between keywords \textit{color\_map} and \textit{perceptual}. Appendix C includes a few more detailed visualizations, from which we can also observe some other potential topics (e.g., \textit{interaction}, \textit{parallel coordinates}, \textit{design}, \textit{3D}, \textit{bubble}, \textit{dimension}, etc.).  





\begin{figure}[t]
    \centering
    \includegraphics[width=\columnwidth]{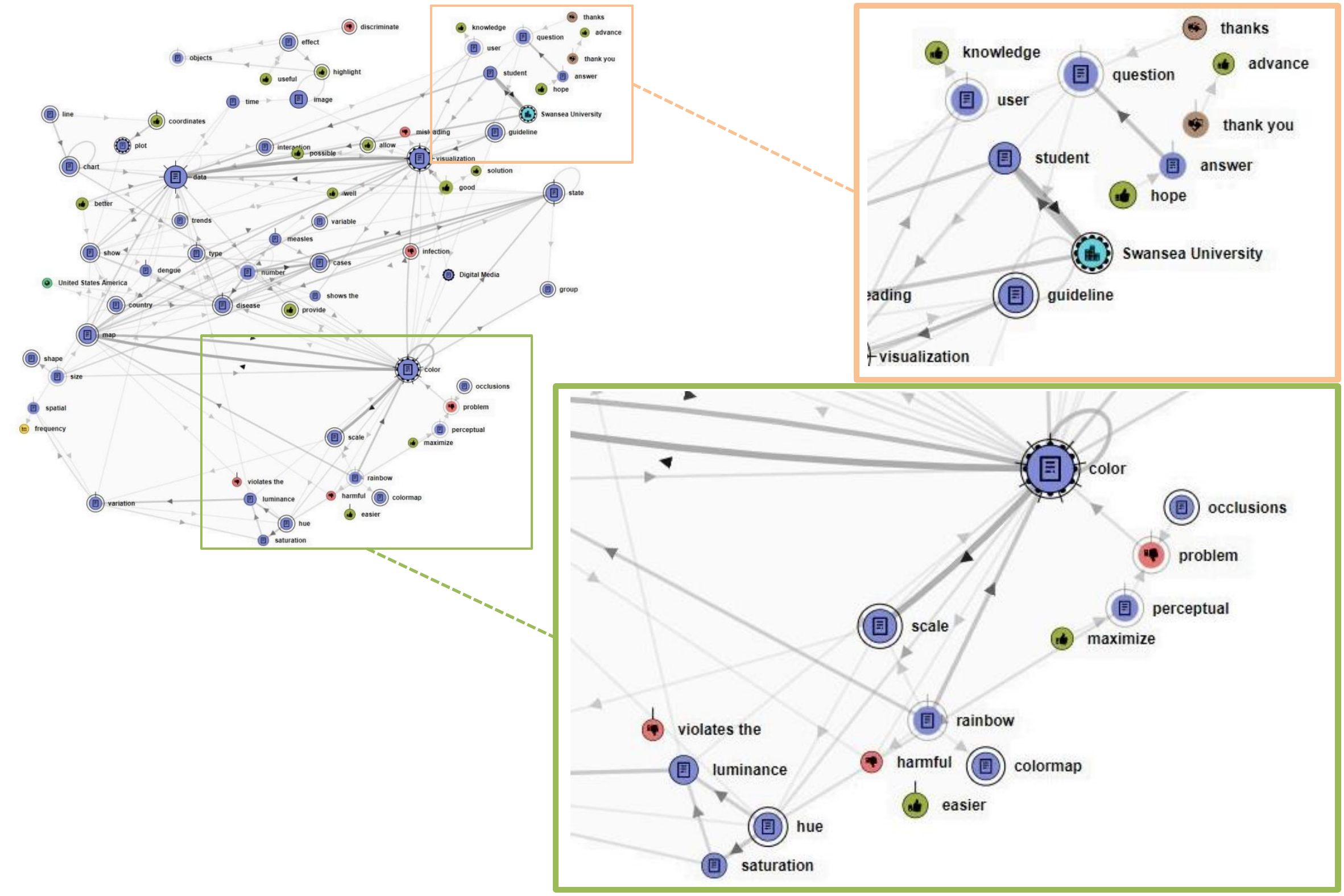}
    \caption{ The Named Entity Graph~\cite{El-Assady2017} depicts the relationship between the most \textit{relevant} entity pairs in a corpus, i.e., entities occurring more often together than an expected threshold. In this graph instance,  the node ``color'' is very prominent, as the use of rainbow colormaps has been extensively discussed in the underlying text.}
    \label{fig:NEREx}
    \vspace{-4mm}
\end{figure}

\begin{figure}
    \centering
    \includegraphics[width=\columnwidth]{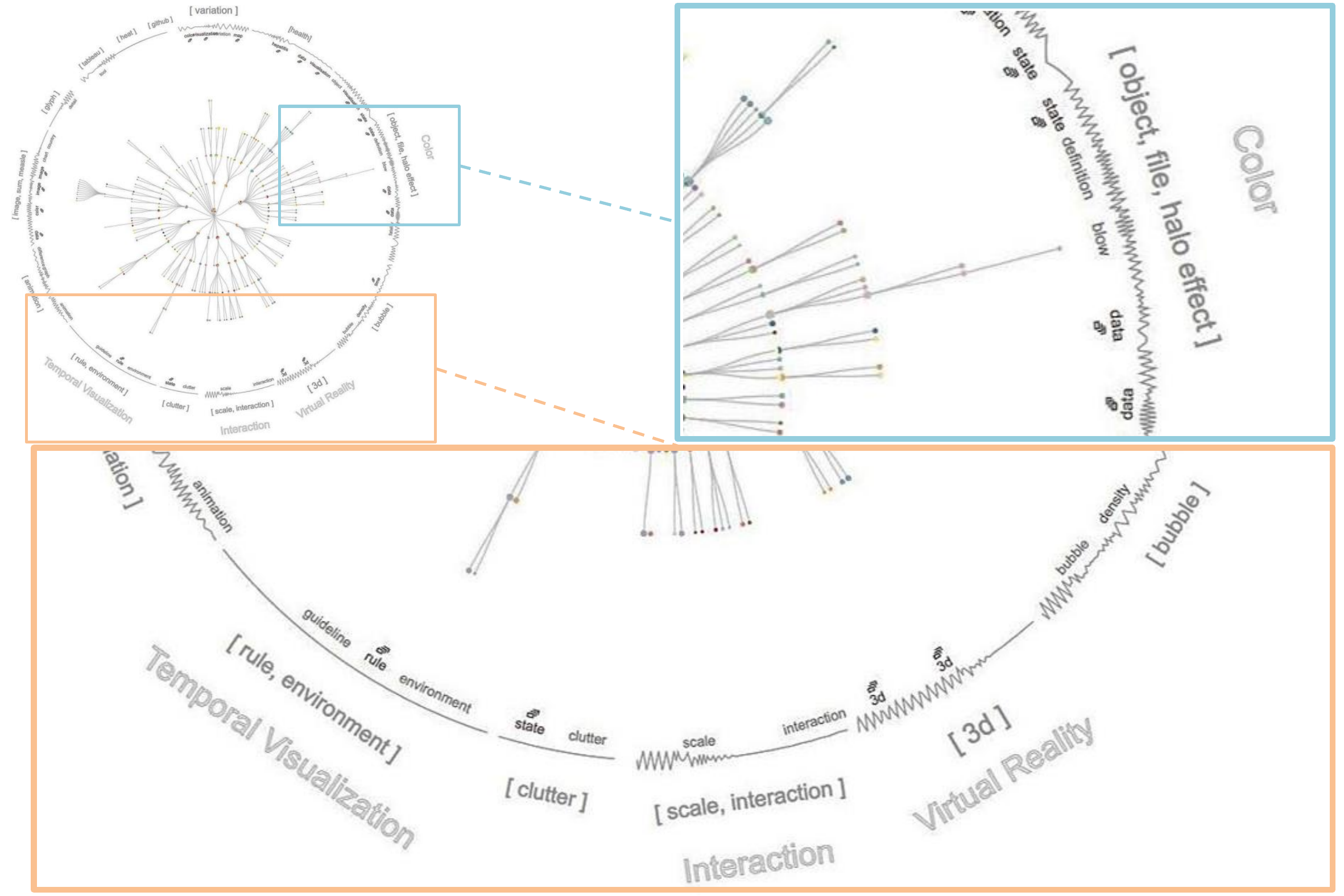}
    \caption{ The  Topic Tree~\cite{el2018visual} \textit{groups all semantically-related utterances} (as leaf nodes) in a hierarchical tree structure with inner-nodes constituting topics and sub-topics. The \textit{outer border} of the tree depicts the \textit{top-descriptors} of each topic branch and the \textit{uncertainty of the respective utterance assignment} to that topic. }
    \label{fig:topics}
    \vspace{-4mm}
\end{figure}

The second technique we used is the \textit{Named Entity Graph}~\cite{El-Assady2017}, as shown in \autoref{fig:NEREx}. This visualization approach is based on named-entity recognition~\cite{manning2014stanford}, dividing descriptive words into ten categories. It considers entities to be related, if they occur within the same sentence in a window of a given size (here: five). The number of times two entities occur together in a corpus determines their edge weight. Using a force-directed layout, entities (nodes) are placed on a canvas, connected by directed edges (showing the order in which the entity pair appeared in the text). Nodes are additionally color-coded to show their category, e.g., purple nodes for content keywords, blue nodes for organizations, orange nodes for person names, and green and red nodes for positive and negative sentiment respectively. The size of each node encodes the number of times this entity appears in the text. In the example shown in \autoref{fig:NEREx}, the above-mentioned discussion on \textit{colormaps} is again prominent in the lower-right corner of the graph, accompanied by the keywords \textit{rainbow}, \textit{easier}, and \textit{harmful}. In the upper-right corner, conversational keywords are connected with the entity ``\textit{Swansea University}'', referring to the 
assignment described in Section \ref{sec:swansea}.

The third technique we applied is \textit{Incremental Hierarchical Topic Modeling}~\cite{El-Assady2018ProgressiveFramework} to extract the \textit{Topic Tree} shown in \autoref{fig:topics}. This technique groups all posts (as leaf nodes) in a hierarchical tree structures (with all inner nodes considered topics and sub-topics).  On the outer edge of the tree, the \textit{topic descriptors} for the level-1 branches are shown. Using a technique called \textit{topic backbone}~\cite{El-Assady2018ProgressiveFramework} the topic tree can be primed towards expected clusters. If found, these are shown as the outermost keywords, e.g., \textit{Color}, \textit{Temporal Visualization}, etc. Other clusters found in the topic tree include topics on \textit{Interaction} and \textit{Virtual Reality}, as well as a large group on a discourse on \textit{health} data. In addition, smaller topics such as one on commercial visualization software (keyword: \textit{tableau}) have been extracted as thematically separable from the rest of the topics in the tree.


Using text analysis as an additional means of open coding has complemented the manual GT processes reported in Section \ref{sec:VisGuides}, while text visualization has enabled humans to connect the keywords and relationships identified by algorithms with the insights obtained from close reading. In particular, from text visualization, we have made the following observations about the research question \textbf{Q3.}
\begin{itemize}
  \vspace{-1mm}
  \item We have identified a number of keywords corresponding to specific visual representations such as \emph{parallel coordinates} and \emph{treemap}. This suggests that they provide visualization guidelines with one of the most important contexts, and it is desirable to define topics or sub-topics based on such contexts.
  \vspace{-1mm}
  \item Some keywords, such as \emph{color}, \emph{shape}, \emph{space}, \emph{time}, and \emph{perception} are generic to many posts, which indicate some common problems and solutions to be addressed by guidelines. These keywords may define another variable for categorizing visualization guidelines. 
  \vspace{-1mm}
  \item The relationships among various keywords (as shown in Figs. \ref{fig:LEP}, \ref{fig:NEREx}, and \ref{fig:topics}) are prominent in seeking and offering advice with or without using guidelines, and relationships between different concepts should ideally be featured in a taxonomy of visualization guidelines. This finding echoes a similar conclusion based on the language analysis of guideline expressions in \cite{kandogan2016grounded}.
  \vspace{-1mm}
  \item In addition to the individual posts and threads, more information may potentially be extracted from conversational fragments, cross-talks, and commonplaces among threads. Therefore, further text analysis and visualization, e.g., using the \textit{VisArgue platform}~\cite{El-Assady2017c}, may enable us to make new discoveries.
  \vspace{-1mm}
  \item The keywords and relationships identified from VisGuides data sets will continue to evolve in the coming years. They are useful for prompting humans to consider other keywords and relationships that may be featured in different contexts, e.g., publication and education (see also Section \ref{sec:Teaching}).
\end{itemize}

%% file: sections/Swansea.tex
\begin{figure}[t]
   \centering
   \begin{tabular}{@{}c@{\hspace{4mm}}c@{}}
    \includegraphics[width=\linewidth]{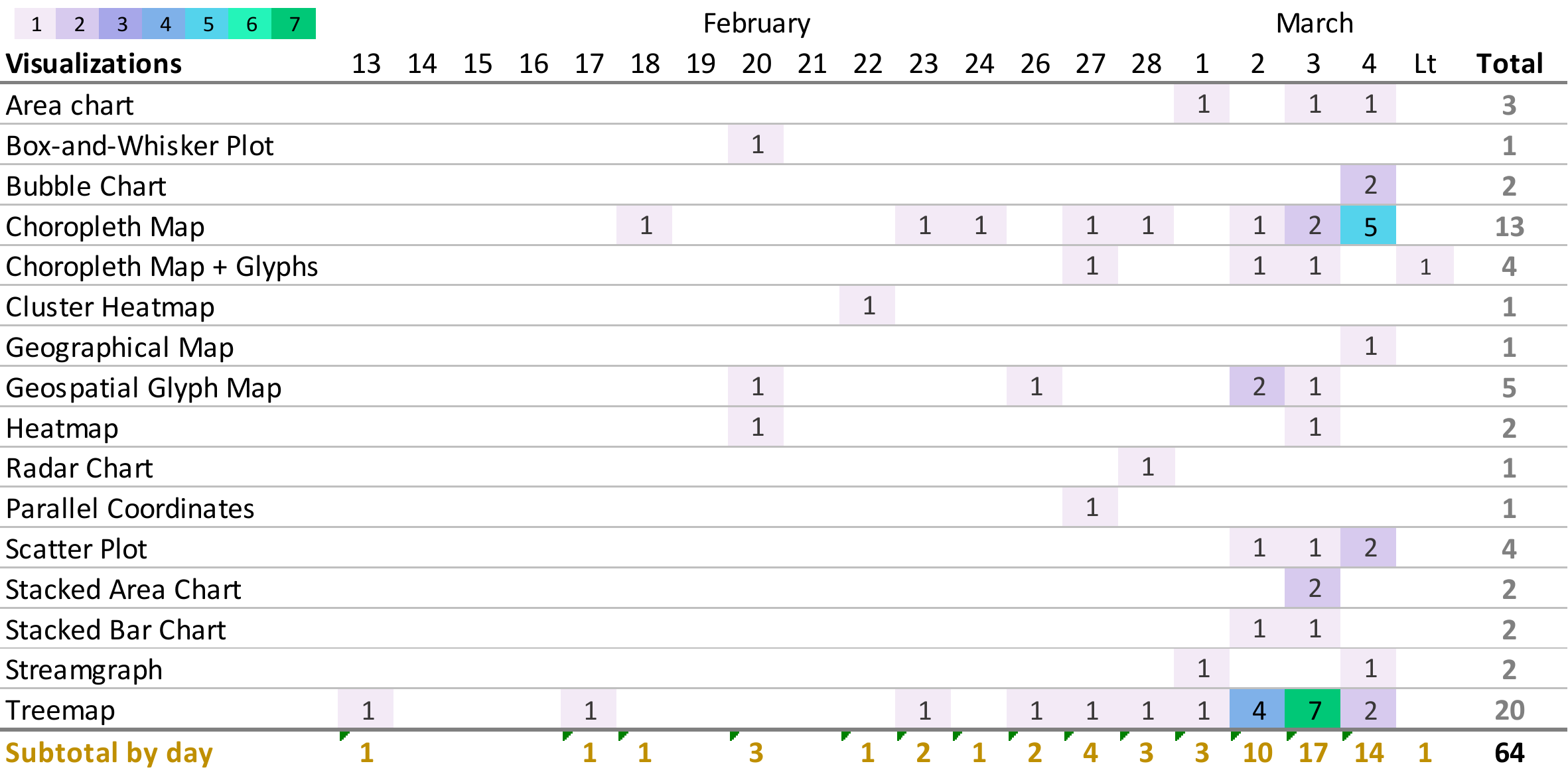} 
  \end{tabular}
  \caption{Activities of Swansea student activity at VisGuides.
  The number in each colored cell is the number of submissions featured a particular type of visualization ($y$) on that day ($x$). The subtotals indicate the popularity of Choropleth Map and treemap.}
  \label{fig:Swansea}
  \vspace{-4mm}
\end{figure}

\section{Organized Discourses in the Context of Education}
\label{sec:Teaching}
%
%
VisGuides was used to support a Data Visualization course in University of Swansea during the academic year 2018/2019. The students were either in their final year of a BSc program or an MSc program in computer science. This created a collection of discourses in the context of visualization education, resulting in a data set $D_{\text{Swansea2019}} \subset D_{20190313}$. 
As mentioned in the previous section, $D_{\text{Swansea2019}}$ exhibits noticeable patterns in text visualization.
As an important aspect of grounded theory is to look systematically for theoretical sensitivity in terms of contexts where data is collected, a hypothesis is proposed, and a conclusion emerges \cite{Strauss:1998:book}.
We organized a parallel grounded theory process for analyzing $D_{\text{Swansea2019}}$, focusing on the \textbf{Q4} and \textbf{Q5} in Section \ref{sec:Introduction}, with further consideration of \textbf{Q1-Q3}.  


%
%
%
%

\paragraph{\textbf{The Assignment.}}
\label{sec:swansea}
The students were encouraged to use VisGuides to seek advice in order to improve their 
visualizations created as part of an assignment. The students were required to select 
epidemiological data sets from the repository of Project Tycho~\cite{burke:project}, 
and to create visualizations to enable the understanding of the data. Students were 
expected to work on spatio-temporal data sets, and use visual representations more 
complex than basic statistical charts.

As an optional component of the assignment, students were instructed to ``request 
professional help from VisGuides.org.'' The task was to choose a created image, upload 
it with an appropriate description, and pose a question(s).
The task of creating visualization was mandatory while that of using VisGuides was not. 
The detailed specification of the assignment is included in the supplementary materials.


The assignment was set on February 4, 2019. A total of 64 students out of 71 participated in the VisGuides aspect of the assignment.
From Fig. \ref{fig:Swansea}, we can observe that 41 students submitted their posts for seeking advice within the last three days of the assignment deadline (11am on March 4). The majority of the posts received at least one reply. One student commented: 
%

``\textit{Personally the question I myself asked did not improve much directly for my assignment, 
however this may be [because] I had already learnt a great deal from reading all the questions and replies
made by other students. I also learnt quite a bit about visualisation techniques I did not even 
consider for my own assignment through VisGuides.
At first I did not think this part would contain much substance 
 as I guessed all questions would be very similar, however, I now  think it was probably the most valuable aspect of the 
coursework.}''
This summarizes the typical learning experience gained by many students, in terms of \textbf{Q4} (Did VisGuides supplement and enhance the learning experience of those who studying the topic of data visualization?).

\paragraph{\textbf{Analysis of $D_{\text{Swansea2019}}$ and Our Findings.}}
As shown in Fig. \ref{fig:Swansea}, 17 different chart types were featured in the students' posts. Among them, 17 students selected choropleth maps (with out without glyphs) and 20 selected treemaps as their visual designs.
In terms of \textbf{Q5} (What visualization guidelines mentioned could be included in the lectures on data visualization?), close-reading of students' posts, there were many questions about how to add temporal data to geo-spatial visualizations.
This indicates that students could benefit from more guidelines on this topic, 
As a result, the content of the data visualization course teaching content is being updated to address this need.

Another recurring theme was the aspect of hierarchical data.  
Many students posed questions about how to construct
hierarchies that would inform trends in the Project Tycho data.
From the questions and resulting replies, it is evident that many 
students struggled with the hierarchical aspect of treemaps.
Again students could benefit from more guidelines on  how to order variables sensibly in the hierarchy of a treemap. Although this was discussed briefly in the class, there is a clear need for dedicating more time to this topic in the future.
Another recurring theme was the treemap layout.
Our close-reading revealed that students generally struggled with the node placement
strategy of treemap layout algorithms.
Students could benefit from guidelines on using treemap layout algorithms.

%% file: sections/KCL-interaction.tex
\section{Empirical Studies}
\label{sec:EmpiricalStudies}
The discussions on VisGuides often suggested that further data collection would be necessary in order for some visualization guidelines to be adequately grounded on data.
In this section, we report two empirical studies for such grounded theory processes.

\begin{figure}[t]
\centering
  \begin{tabular}{@{}c@{\hspace{2mm}}c@{}}
  \includegraphics[height=28.5mm]{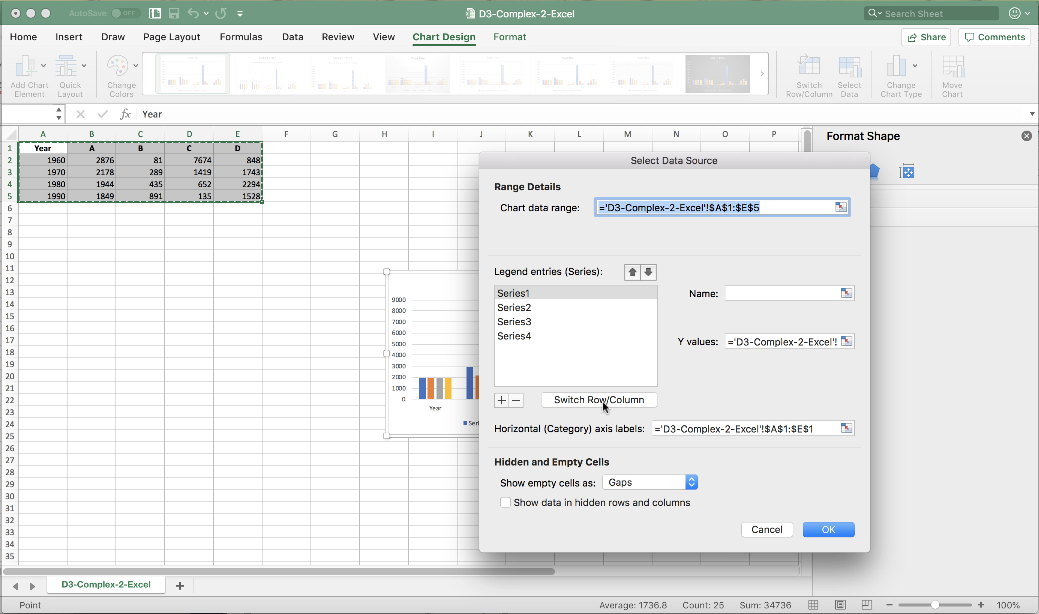} &
  \includegraphics[height=28.5mm]{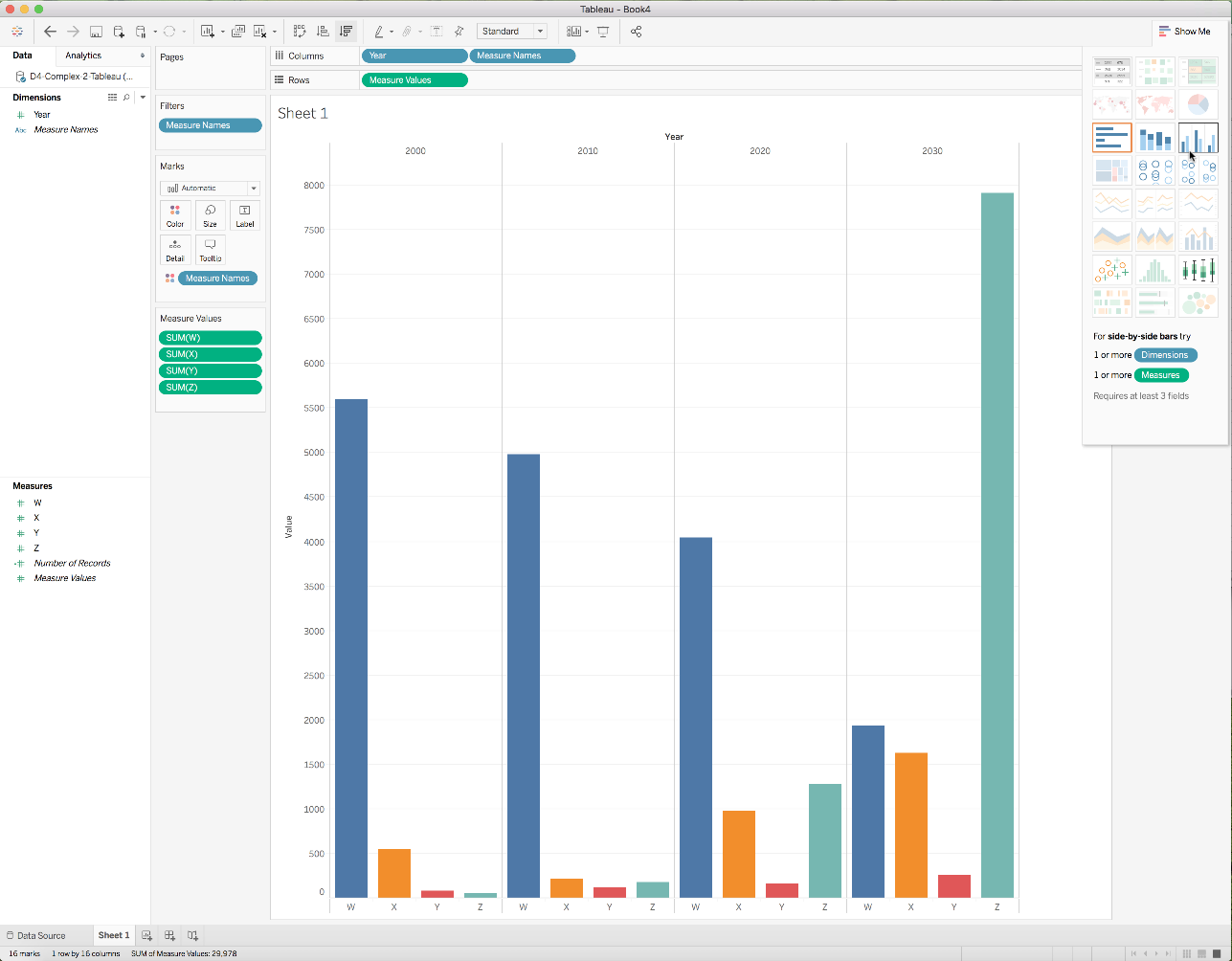}\\
   \small{(a) Using MS-Excel} &
  \small{(b) Using Tableau}
  \end{tabular}
  \caption{Two example screens captured by using Camtasia in an empirical study on the cost of interaction in creating bar charts.}
  \label{fig:KCL-stimuli}
  \vspace{-4mm}
\end{figure}

\begin{figure*}[t]
  \centering
  \begin{tabular}{@{}c@{\hspace{6mm}}c@{\hspace{6mm}}c@{}}
  \includegraphics[height=36mm]{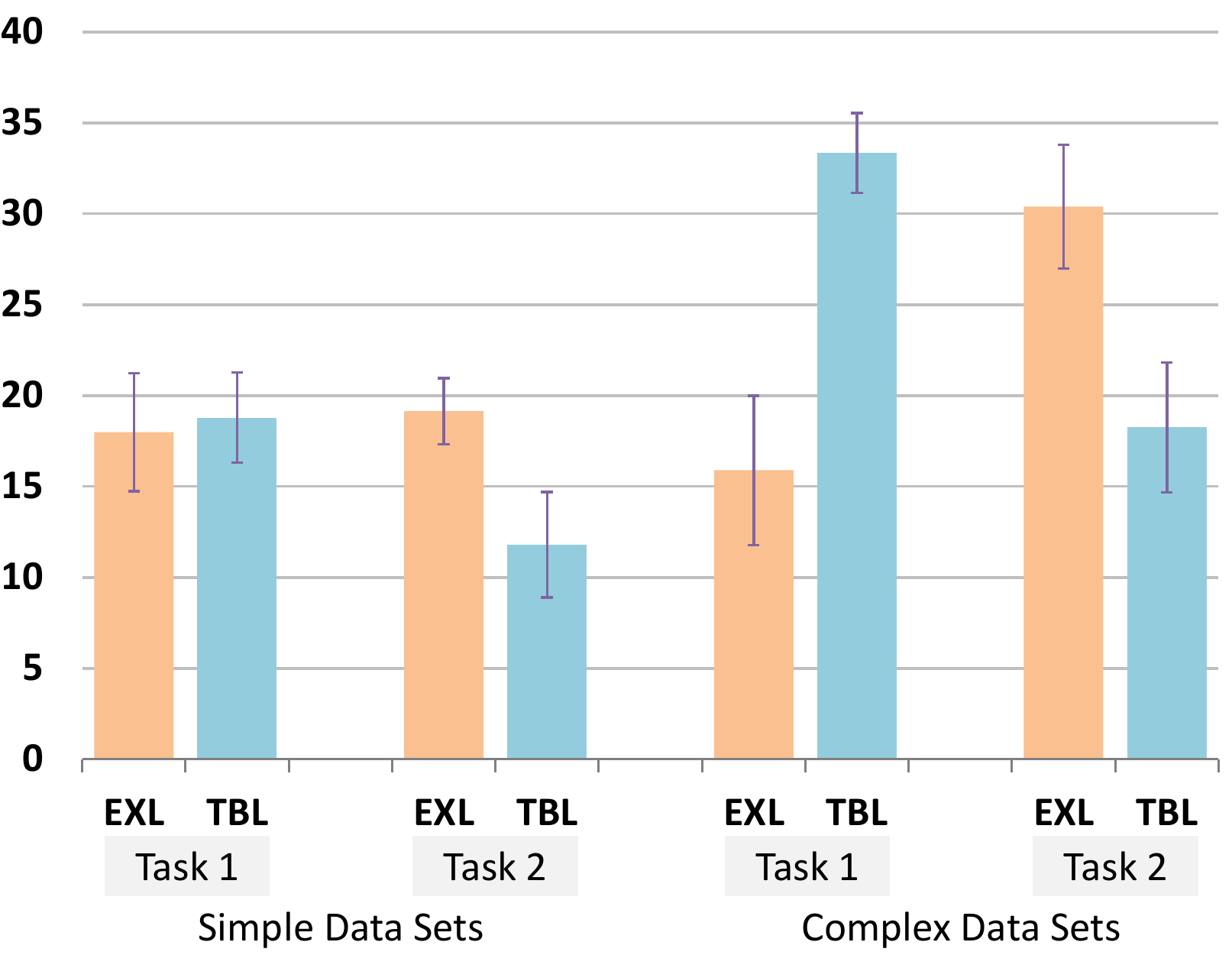} &
  \includegraphics[height=36mm]{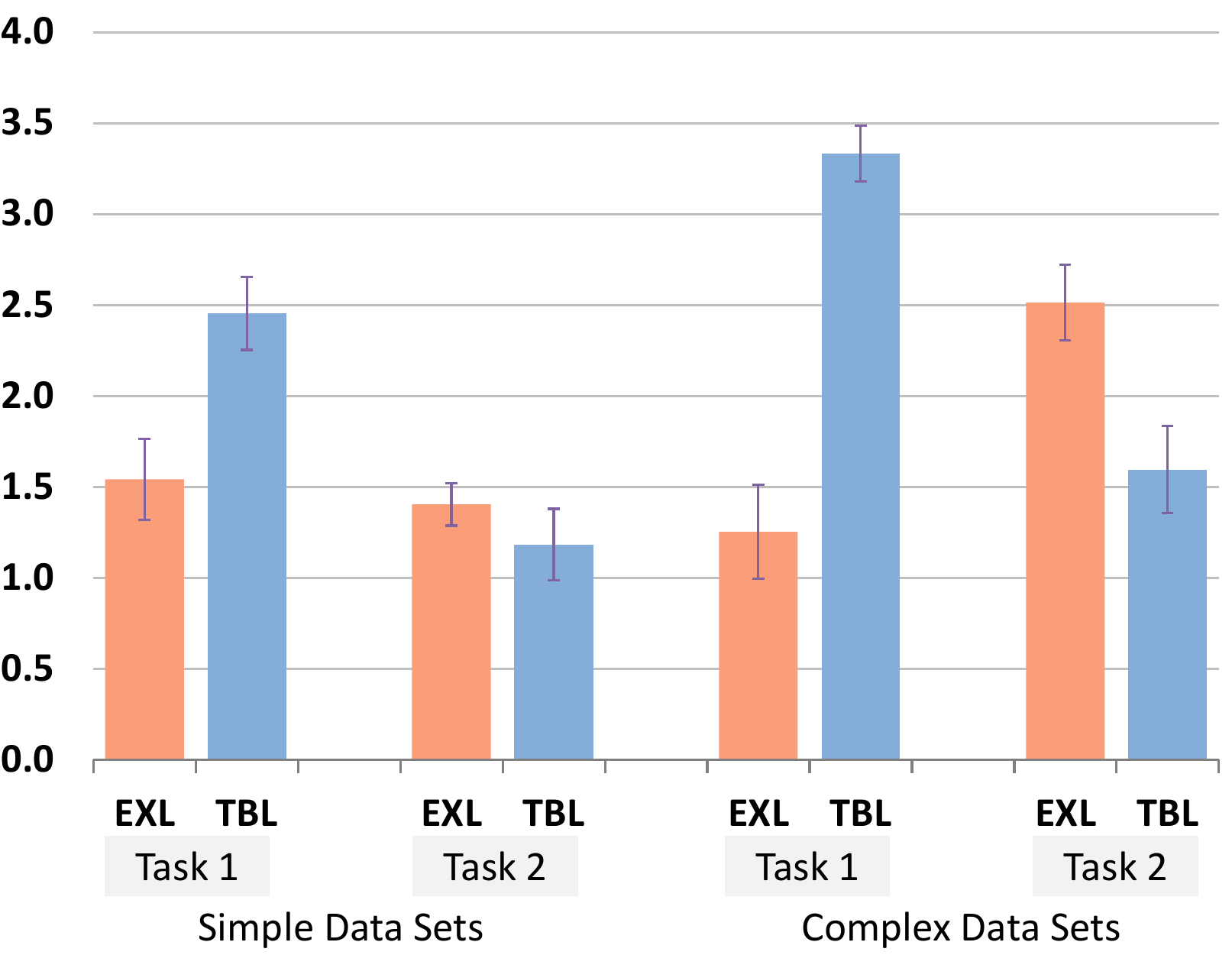} &
   \includegraphics[height=36mm]{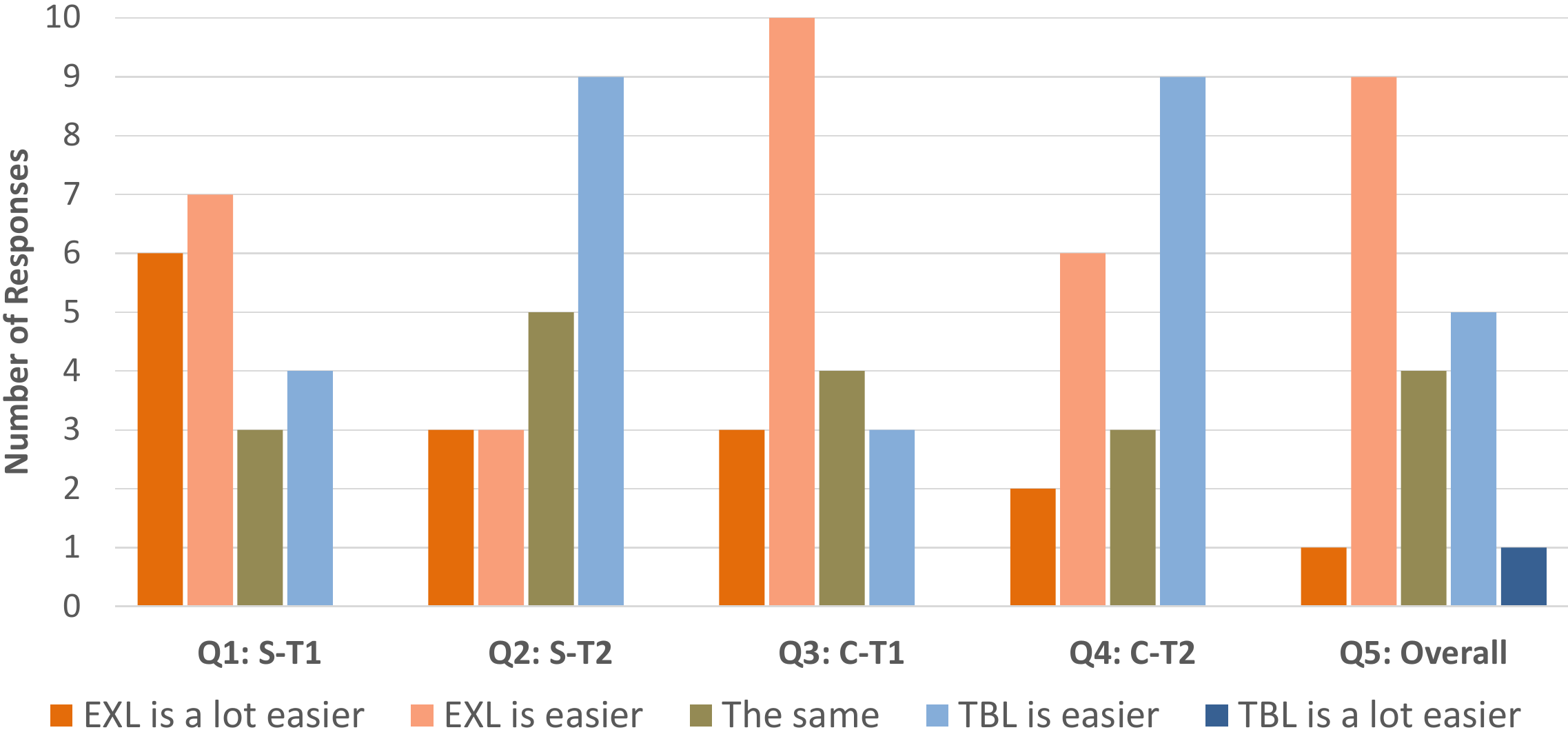} \\
  \small{(a) Average number of interactions (NI)} &
  \small{(b) Average completion time (CT)} &
  \small{(c) Subjective judgments on easiness}
  \end{tabular}
  \caption{An empirical study on the cost of interaction. The data collected indicates that the participants' opinions about the relative easiness of use depend mainly on the task completion time, partly on the number of interactions, and in some cases, on their familiarity of the software.}
  \label{fig:KCL}
  \vspace{-4mm}
\end{figure*}


\subsection{A Study Inspired by ``Action without Interaction''}
\label{sec:KCL}
In September 2017, \textbf{east101} initiated a new thread ``Action without Interaction'' \cite{east101:2017:VisGuides} on VisGuides to discuss a guideline ``Use interaction in visualization sparsely and cautiously ...'' proposed by Gr\"{o}ller \cite{Groller:2008:Dagstuhl}. It received two replies. One (\textbf{groeller}) explained: ``interaction typically requires considerable effort and cognitive load.'' Another (\textbf{jamesscottbrown}) presented a different argument: ``... there is generally little cost (to the user) associated with adding such interactivity to a visualisation ...'' 

We wondered what would be users' opinions if they were asked. Since empirical studies are integral part of grounded theory \cite{Willig:2013:book}, we conducted a controlled experiment to collect data reflecting such ``opinions''.    
We hypothesized that \emph{the amount of interaction may influence a person's opinion as to how easy it is to use a visualization tool}.
If the same task is performed using two different tools, the tool that requires fewer interactions or less interactive time corresponds to ``using interaction'' more ``sparsely and cautiously''.

\vspace{-1mm}
\paragraph{\textbf{Study Design.}}
The study consists of four trials. For each trial, participants are required to create two bar charts to visualize a given data set. Two of the four data sets, Sa and Sb, have two columns of data and are characterized as \emph{simple data sets}. The other two data sets, Ca and Cb, have five columns of data, and are characterized as \emph{complex data sets}.
For each data, participants are first asked to perform a \emph{primary task} (T1) such that a bar chart can be produced directly based on the given data without any manipulation.
This is followed by a \emph{secondary task} (T2) that requires some simple data manipulation before a chart can be generated. The main \emph{independent variable} encodes the variation of two visualization tools used for performing these tasks. We chose to use \emph{MS-Excel} (EXL) and \emph{Tableau} (TBL).
Fig.~\ref{fig:KCL-stimuli} show two screenshots of the two visualization tools respectively.

There are 16 tool-data-task combinations, $\{$EXL, TBL$\}$ $\times$ $\{$Sa, Sb, Ca, Cb$\}$ $\times$ $\{$T1, T2$\}$. To avoid any learning effect when applying different tools to the same data set, participants were divided into two groups randomly. One group were given trials [EXL-Sa, TBL-Sb, TBL-Ca, EXL-Cb] and another were given trials [TBL-Sa, EXL-Sb, EXL-Ca, TBL-Cb]. The data sets Sa and Sb have the same numbers of rows and columns, and they were design to cause little confounding effect when they were used with different tools.
Similarly the data sets Ca and Cb were also designed to cause little confounding effect.
In other words, this allowed us to compare between the two trials in the following pairs (EXL-Sa, TBL-Sb), (TBL-Sa, EXL-Sb), (EXL-Ca, TBL-Cb), and (TBL-Ca, EXL-Cb).

The study focused on two \emph{dependent variables} -- the \emph{number of interactions} and the \emph{task completion time}. Because all participants can access a help-sheet with step-by-step instructions for each task (after 1 minute for Sa and Sb and 2 minutes for Ca and Cb), they were all able to complete the tasks. The accuracy is thus not the focus of this study.


\vspace{-1mm}
\paragraph{\textbf{Procedure.}} Appendix E.1 gives more detailed description about the apparatus used and the procedure for conducting the experiment.

%
For each participant, the experiment consisted of four trials, either [EXL-Sa, TBL-Sb, TBL-Ca, EXL-Cb] or [TBL-Sa, EXL-Sb, EXL-Ca, TBL-Cb]. In each trial, the data set was pre-loaded onto the visualization tools. Participants were given written specification of the tasks together with examples of the bar charts to be created. After an initial attempt of 1 minute for tasks associated with data sets Sa and Sb and 2 minutes for tasks with data sets Ca and Cb, participants were allowed to make use of a help-sheet to complete the task. At the end of the second trial, participants ware presented with two questions (for T1 and T2 with Sa and Sb) about the easiness of using the two visualization tools. They are asked to rank the relative easiness using the Likert scale. Similarly at the end of the fourth trial, participants were presented with three questions (for T1 and T2 with Ca and Cb, and for the overall impression).

\vspace{-1mm}
\paragraph{\textbf{Participants.}}
A total of 20 participants (12 females, 8 males) took part in this experiment in return for a \pounds5 Amazon voucher. Participants were recruited from King's College London. They belong to both the student and working communities and are from a number of disciplines, including history, psychology, business, politics, mathematics, law, and computer science. 17 participants were in the 20-29 age group and 3 were in the 18-19 age group. In rating their experience in using MS-Excel, 3 rated ``beginner'', 16 rated ``intermediate'', and 1 rated ``advanced''. In the rating of their experience in using Tableau, 18 rated ``never'', 1 rated ``beginner'', and 1 rated ``intermediate''.

\vspace{-1mm}
\paragraph{\textbf{Results and Discussion.}}
From the collected data, we first extracted the number of interactions (NI) and the task completion time (CT) per participant, trial and task, yielding $20 \times 4 \times 2$ pairs of (NI, CT). After confirming that there were no confounding effects between Sa and Sb and between Ca and Cb, we created a combined group EXL-S-T1 by merging the data of EXL-Sa-T1 and EXL-Sb-T1, and the other seven combined groups in a similar manner.
We therefore compute eight pairs of average (NI, CT) for EXL-S-T1, TBL-S-T1, EXL-S-T2, TBL-S-T2, EXL-C-T1, TBL-C-T1, EXL-C-T2, and TBL-C-T2.

Figs.~\ref{fig:KCL}(a,b) show the eight average values of NI and eight average values of CT respectively.
Fig.~\ref{fig:KCL}(c) summarizes the participants' answers to the five questions. Each bar indicates the number of participants who rated the easiness in one of the five Likert points.
The ANOVA results indicate that most pairwise comparisons in Figs.~\ref{fig:KCL}(a,b) are significant, except the first two NI averages in Fig.~\ref{fig:KCL}(a) for EXL-S-T1 vs. TBL-S-T1. The initial analysis also cast a doubt on the statistical significance of the pairwise comparison EXL-S-T2 vs. TBL-S-T2 in terms of both NI and CT. A close examination of the data revealed a single outlier because a participant resisted to consult the help-sheet after many failed attempts in this particular trial. After removing the data of this outlier, the ANOVA indicated convincing statistical significance. 
From these statistics, we can make the following observations:
%
\begin{itemize}
  \vspace{-1mm}%
  \item For the simple data sets (Sa and Sb), MS-Excel and Tableau incurred similar numbers of interactions for performing the primary task (T1), while MS-Excel required less time than Tableau (which could be influenced by familiarity). The participants' opinions on the relative easiness are in favor of Excel.
  \vspace{-1mm}%
  \item For the simple data sets (Sa and Sb), Tableau incurred fewer interactions and less task completion time for performing the secondary task (T2). The participants' opinions are in favor of Tableau.
  \vspace{-1mm}
  \item For the complex data sets (Ca and Cb), MS-Excel incurred noticeably fewer interactions and shorter task completion time for performing the primary task (T1). The participants' opinions are overwhelmingly in favor of Excel.
  \vspace{-1mm}
  \item For the complex data sets (Ca and Cb), Tableau incurred fewer interactions and less task completion time for performing the secondary task (T2). However, the participants' opinions are marginally in favor of Excel. This inconsistency may be caused by the participants' familiarity of the software or their experience of the proceeding task. 
  \vspace{-1mm}
  \item The participants' overall impression is in favor of MS-Excel, which is consistent with the overall NI and CT measures collected.
\end{itemize}
Based on the above observations, we conclude that participants' opinions on the relative easiness depend mainly on the task completion time, and to some extent also on the number of interactions. In some cases, the familiarity of the software may modify the direct correlations among NI, CT, and subjective judgment.
In term of \textbf{Q6} in Section \ref{sec:Introduction}, while this study provides a new collection of data on which the guideline ``Use interaction in visualization sparsely and cautiously ...'' \cite{Groller:2008:Dagstuhl} can be grounded, we recognize that more grounded theory activities are necessary in order to achieve theoretical saturation.

%% file: sections/Konstanz-VR.tex
\begin{figure}[t]
\includegraphics[width=\columnwidth]{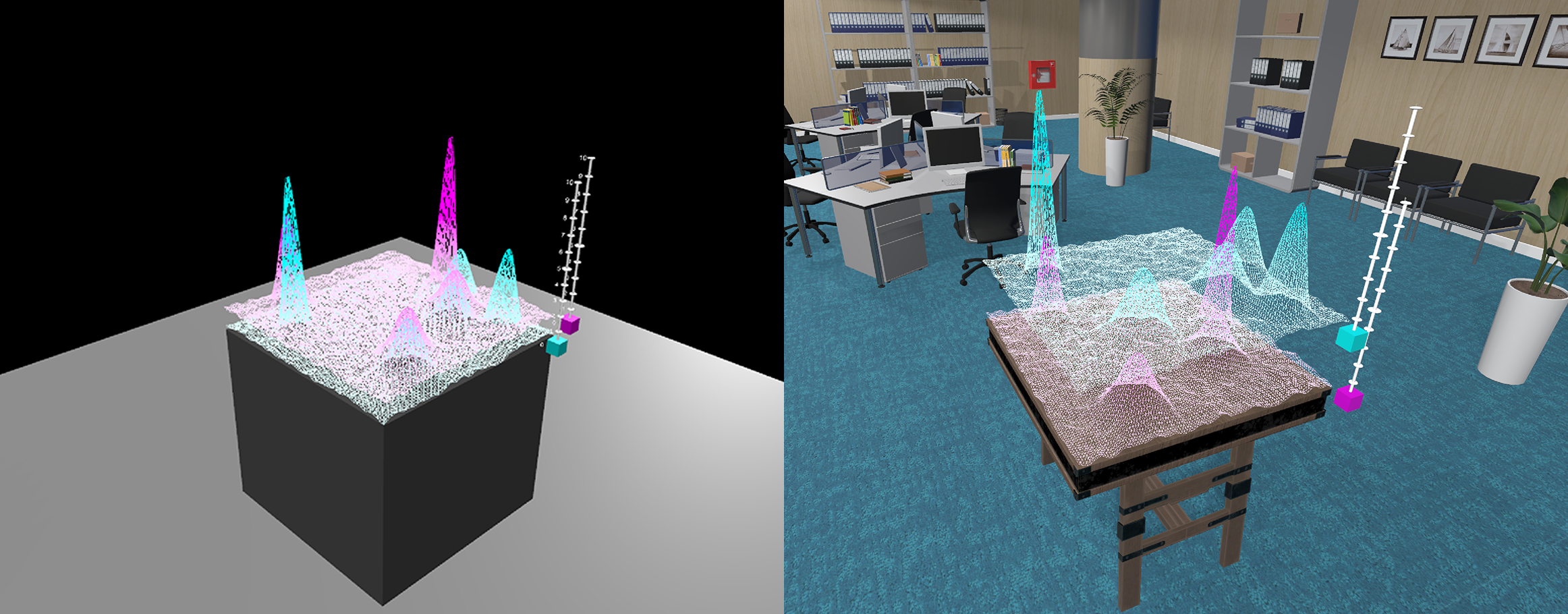}
  \centering
  \caption{Left: Plain condition, Right: Office condition. In each condition, participants performed an comparative visual exploration task. }
 \label{fig:Konstanz-stimuli}
 \vspace{-4mm}
\end{figure}

\begin{figure*}[t]
  \centering
  \begin{tabular}{@{}c@{\hspace{2mm}}c@{\hspace{4mm}}c@{\hspace{2mm}}c@{}}
    \includegraphics[height=44mm]{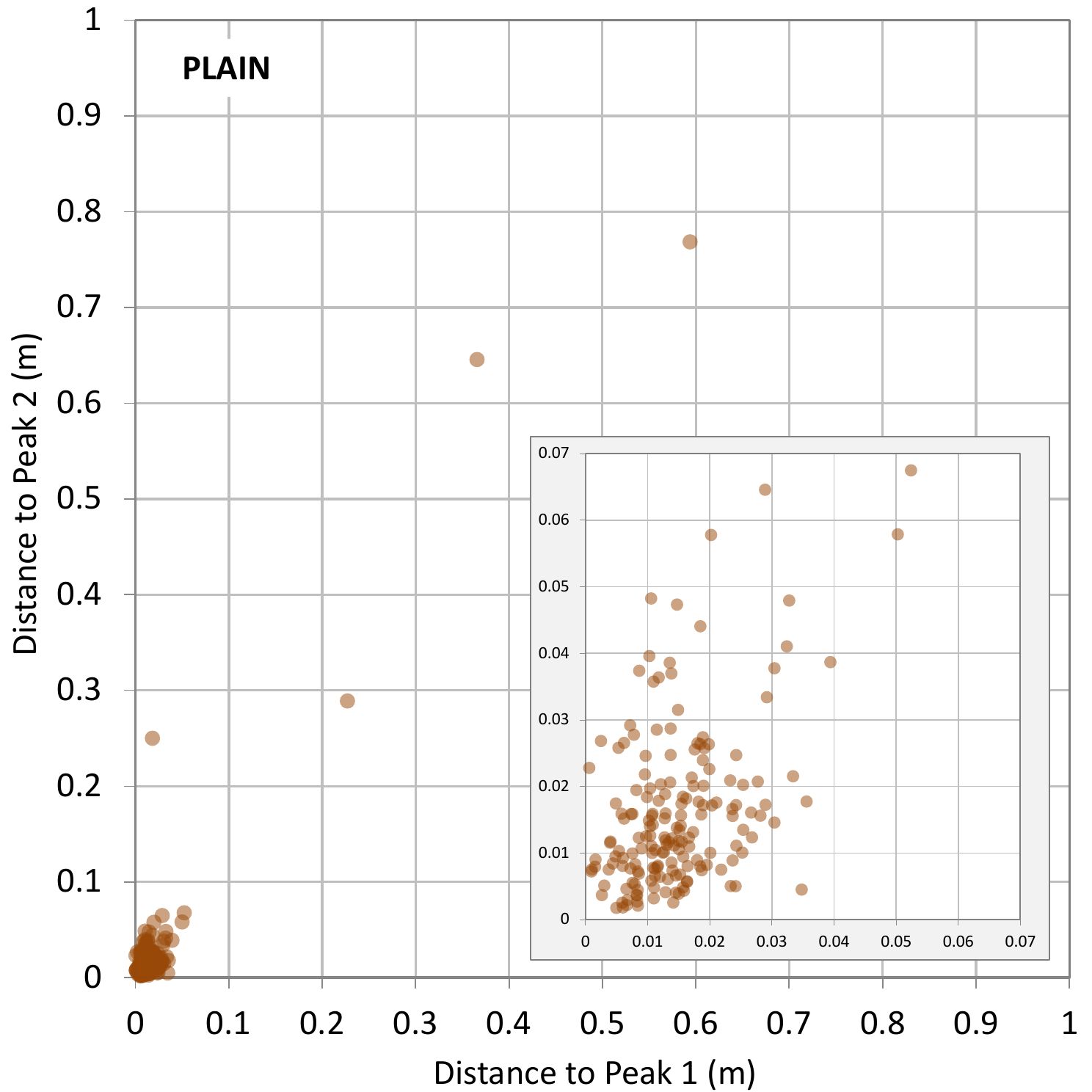} &
    \includegraphics[height=44mm]{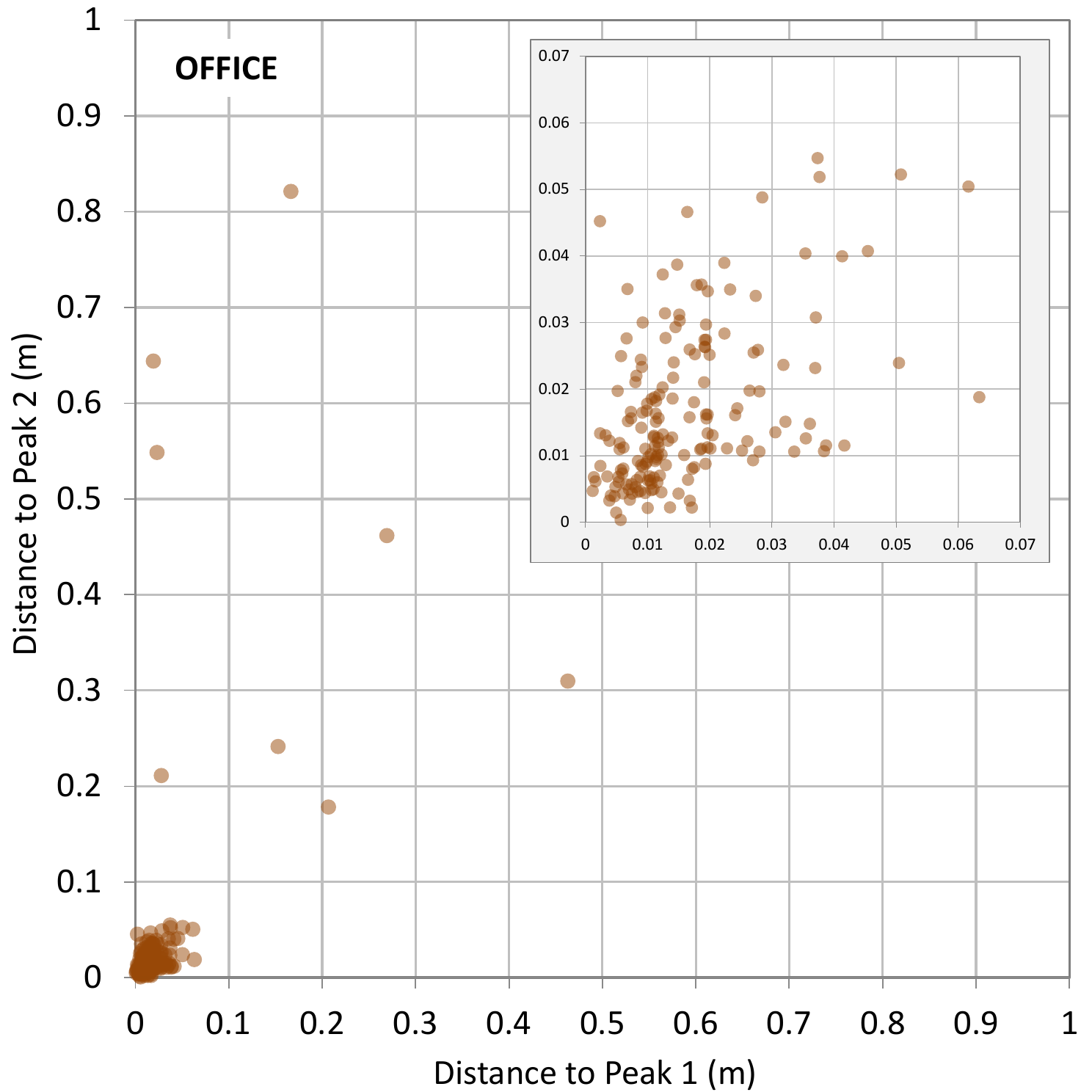} &
    \includegraphics[height=44mm]{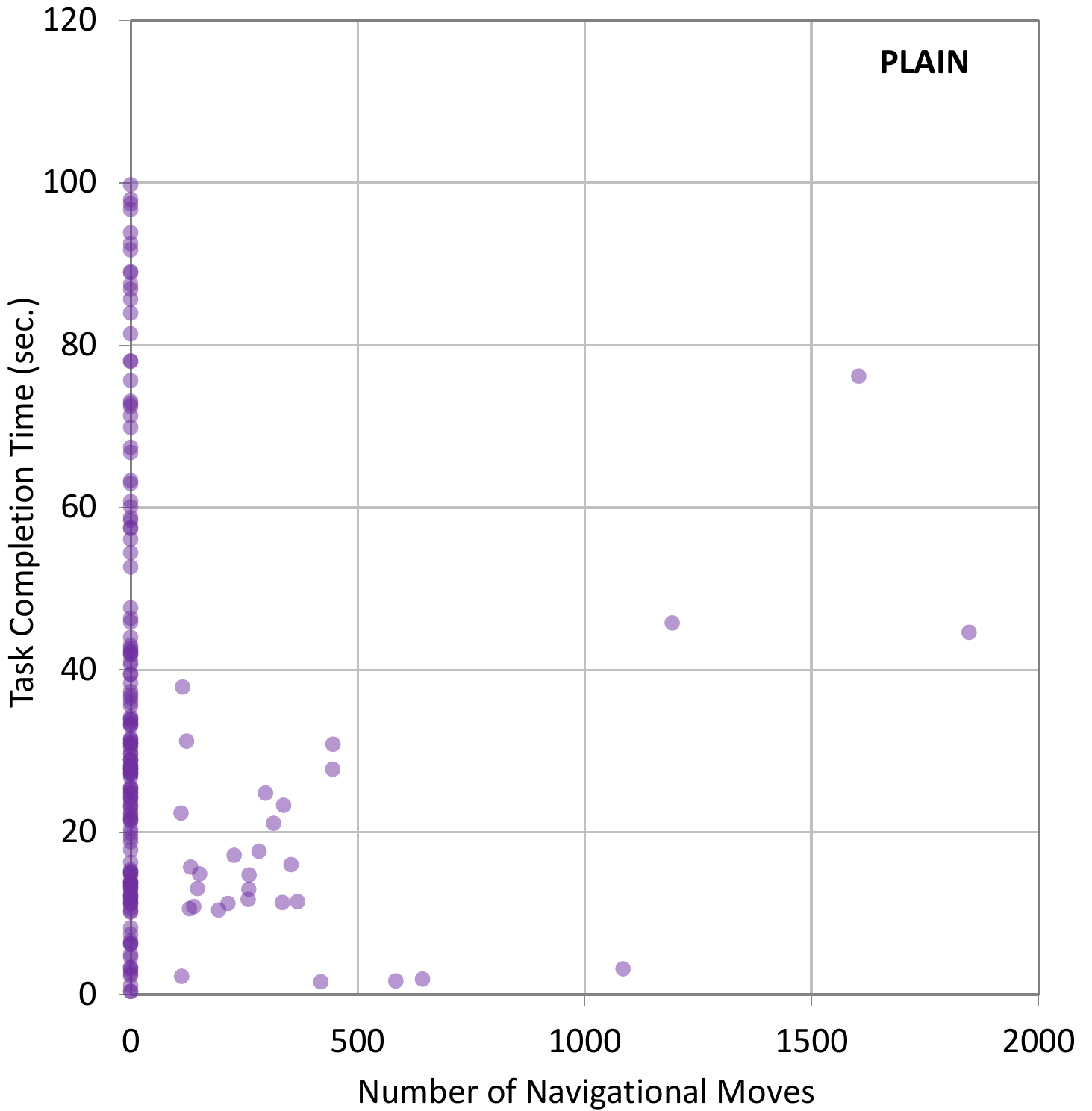} &
    \includegraphics[height=44mm]{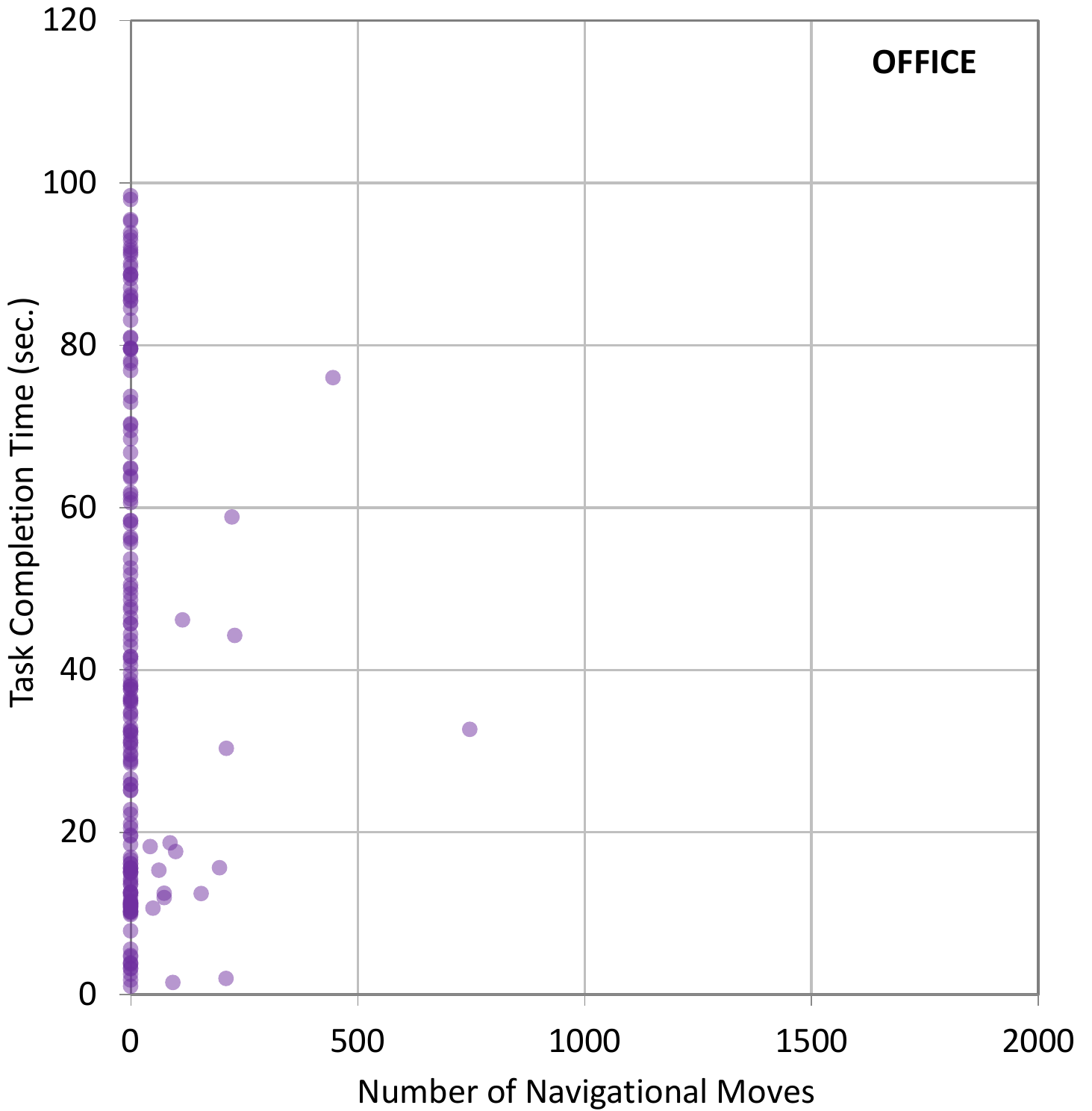} \\
    \small{(a) Plain: distances to Peak 1 and Peak 2} &
    \small{(b) Office: distances to Peak 1 and Peak 2} &
    \small{(c) Plain: \# moves vs. completion time} &
    \small{(d) Office: \# moves vs. completion time}
  \end{tabular}
  \caption{Four scatter plots show the collected measurements from the 180 trials in the plain 3D environment, and the 180 trials in the office VE. The four measurements are the distance from the pointer location to Peak 1 (i.e., the correct peak in surface 1), the distance to Peak 2, the number of layer movements, and the task completion time.}
  \label{fig:Konstanz-ScatterPlots}
  \vspace{-4mm}
\end{figure*}
%
\subsection{A Study Inspired by ``Don't Replicate the Real World''} 
\label{sec:Konstanz}
In May 2018, as part of literature research for a PhD project, \textbf{matthias.kraus} initiated a new thread ``(Don't / Do) Replicate the Real World in VR?'' \cite{kraus:2018:VisGuides}. The discussion focused on the visualization guideline ``Rule \#7: Don't Replicate the Real World'' proposed by Elmqvist in a blog \cite{Elmqvist:2017:blog}. The initial post suggested a hypothesis that replication of real world could create ``familiar surroundings to lower the learning curve'' and ``at least'' would disadvantage task performance. It attracted seven replies, including two replies from Elmqvist (\textbf{elm}).

Following the discussions, the PhD student who initiated the thread decided to conduct a controlled experiment to collect more data in order evaluate the guideline in the context of his hypothesis. This became part of the research question \textbf{Q6} in Section \ref{sec:Introduction}.


\paragraph{\textbf{Study Design.}}
The data sets used in this study are height fields in the form of $y=H(x,z)$, which are commonly used to depict bi-variate data distributions and terrain landscapes. 
As shown in Fig. \ref{fig:Konstanz-stimuli}, the main data sets in each stimulus are two 
height fields, $H_1$ and $H_2$, displayed as two 3D surfaces, one floating on the top of another, in a visually-realistic VE. 

The \emph{independent variable} encodes two scenarios: whether or not the data is rendered as plain 3D graphical objects or a more metaphoric virtual objects in an office environment.  
The \emph{visualization task} is to identify the \emph{common peaks} in both surfaces and to select two highest peaks among all common peaks. Here a common peak is defined as the location $(x, z)$ such that $y_1 = H_1(x, z), y_2 = H_2(x, z)$, $y_1 = y_2$, and $(x, y_1, z)$ and $(x, y_2, z)$ are both local maxima.
The \emph{dependent variables} include the distances from the pointer location to the two highest common peaks ($D_1$, $D_2$), the task completion time ($T$), and the number of layer movements made during each trial ($N_m$). A layer movement required a participant to grab a surface and move it upwards or downwards. When the two surfaces are more separated, one can see the two surfaces more distinctively, and when they are close to each other, one can see the overlapping of the peaks more precisely. 
As the participants were allowed to take as much time as they wish to complete the task, the task completion time is the main focus of the study. 

\begin{figure}[t]
    \centering
    \includegraphics[width=70mm]{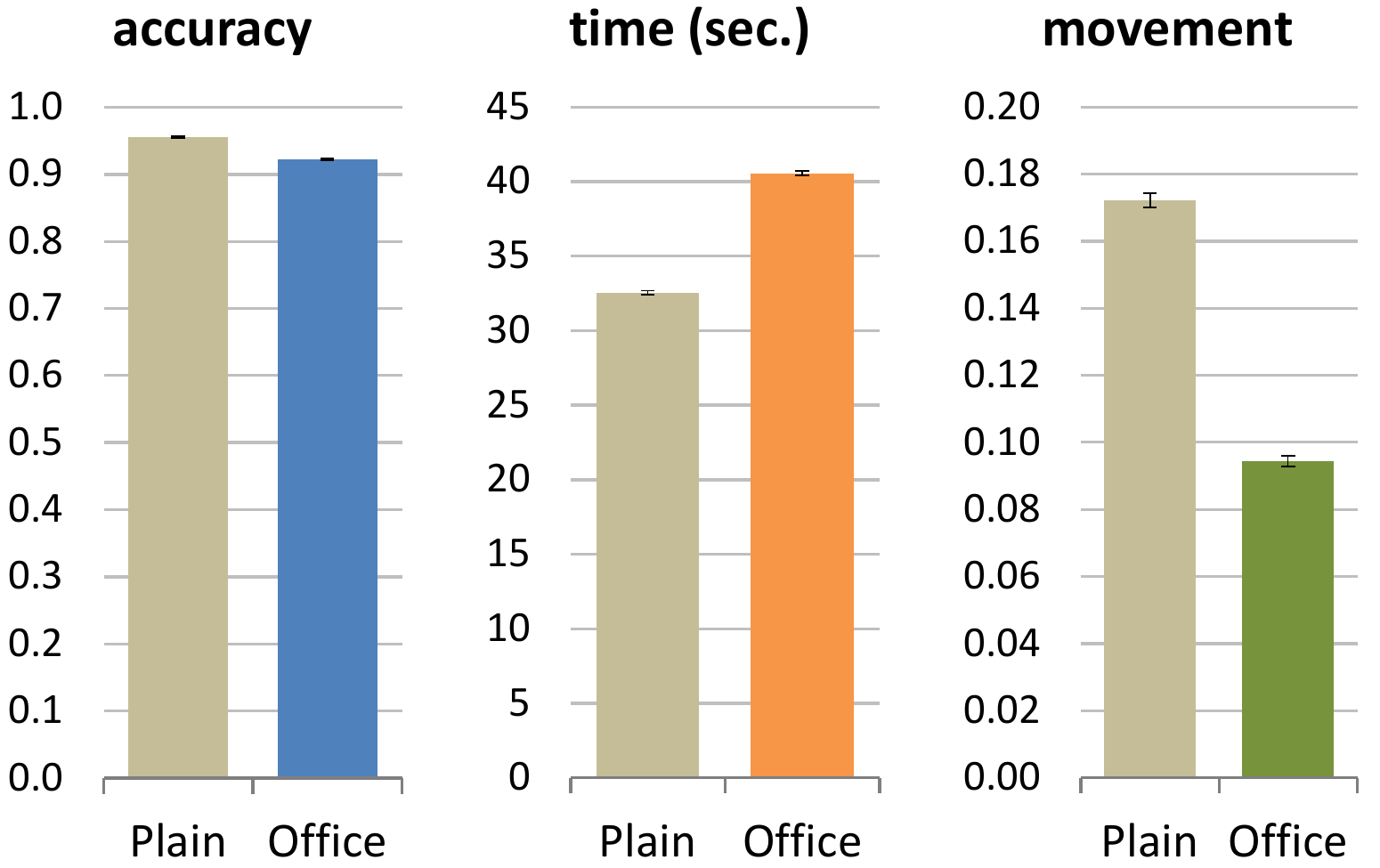}
    \caption{Comparing the average accuracy ($100\%=1$), the average task completion time, and the average percentage of trials with layer movement ($20\%=0.2$) in the plain 3D environment and the office VE.}
    \label{fig:Konstanz-Barcharts}
    \vspace{-4mm}
\end{figure}



\vspace{-1mm}
\paragraph{\textbf{Procedure.}}
Fig.~\ref{fig:teaser}(c) shows the main \emph{apparatus} used  in this experiment. Appendix E.2 describes the apparatus and the procedure in detail.
%
%
Participants are randomly divided into two groups. One group began with a block of 15 trials with the plain 3D environment, and then continued with a block of 15 trials with the office-based VE. Another group began with the office-based VE and continued to the plain 3D environment. The stimuli data sets (i.e., the height fields) used in the 30 trials were randomized.    


\vspace{-1mm}
\paragraph{\textbf{Participants.}}
12 participants (2 females, 10 males) were recruited from the Department of Computer Science, University of Konstanz. All had medium to advanced levels of experience in using VR equipment and performing visualization tasks. As each participant completed 15 trials in the plain 3D environment, and 15 trials in the office VE, there are 180 trials in each of the two conditions.


\paragraph{\textbf{Results and Discussion.}}
For each trial, we measured four dependent variables, including the distances from the pointer location to the two ground-truth common peaks ($D_1$, $D_2$), the task completion time ($T$), and the number of layer movements during the trials ($N_m$). The participants were allowed to take as much time as they wish to complete the task.
Fig. \ref{fig:Konstanz-ScatterPlots} shows these four measurements for the 180 trials in the plain 3D environment and the 180 trials in the office VE.

From Fig.~\ref{fig:Konstanz-ScatterPlots}(a,b), we can observe that most distance measures fall within the $0.07m$ (see the inset). As the base of the height fields is of $1m^2$, participants performed their task with reasonably high accuracy. We can also notice that the data points in (a) are slightly closer to (0, 0) than those in (b). Based on the observation, we define the correct task performance as both distance measures are within 0.05m, i.e., $D_1 \leq 0.05 \, \cap \, D_2 \leq 0.05$. As shown in Fig. \ref{fig:Konstanz-Barcharts}(left), from this definition, we derived the average accuracy for the plain 3D environment as 96\% and that of the office VE as 92\%. However, the two-tailed $t$-test indicates that the variance is statistically not significant.

Fig.~\ref{fig:Konstanz-ScatterPlots}(c,d) shows the number of layer movements vs. the task completion time under the two conditions. There are slightly more movements in the plain environment than the office VE, while most trials do not feature any movements. As shown in Fig.~\ref{fig:Konstanz-Barcharts}(right), about 17\% of the trials in the plain environment are with movement, while there are only 9\% of the trials in the office VE. The two-tailed $t$-test indicates that this variation is statistically significant ($p \approx 0.03$).

Interestingly, movements do not appear to correlate with the task completion time. As shown in Fig. \ref{fig:Konstanz-Barcharts}(middle), the average task completion time for trials in the plain environment is of 32.5 sec. and that in the office VE is of 40.6 sec. The two-tailed $t$-test indicates that this variation is statistically significant ($p < 0.01$). 

Because participants were allowed to take as much time as they wish to complete a task, we anticipated that the accuracy would be similar. We were intrigued by the two opposite indications by the average task completion time and the average layer movement. We considered the following possible explanations:
\begin{itemize}
  \item The office VE have provided more visual cues for 3D perception than the plain 3D environment. When in the plain environment, some participants might have felt a need to move surfaces to enhance 3D perception.
  \vspace{-1mm}
  \item The office background may have caused more distraction to the participants than the plain 3D environment. The judgment on the highest peaks through visual comparison could be affected by such distraction. When in the office VE, participants might have taken more time to ascertain their judgment.
\end{itemize}

In terms of the task completion time, the results of this study provided a piece of supporting evidence for the guideline ``Rule \#7: Don't Replicate the Real World'' \cite{Elmqvist:2017:blog}. In terms of the layer movement, the results of the study suggest that the visually-embellished office background may help reduce the need for moving surfaces, which is a type of interaction and incurs costs (see also the previous section). Meanwhile, we can also observe that the cost of interaction is not always directly translated to the cost of task completion time. Clearly, according to grounded theory, further data collection is required to study both guidelines discussed in this section.


%% file: sections/appendix-A.tex
\section*{\textbf{A. Variables and Categories}}
\label{app:Categories}
This appendix contains the specification of the categorization scheme used in this paper. The part of the scheme for human coding was obtained after a number of iterative GT processes. Following the GT principles, the categorization scheme will continue to evolve until it reaches theoretical saturation.

\subsection*{A.1 Terminology and Criteria}
The following terminology is used in this work. 
\begin{itemize}
  \vspace{-1mm}
  \item \textbf{Variable} --- a variable defines a \emph{concept} (e.g., color) that may have different \emph{states} (red, yellow, green, etc). In a different context, it is often referred to as \emph{variable name} (when a variable is confusingly meant as a value), \emph{alphabet} (in information theory), \emph{dimension} (in dimensionality reduction), \emph{factor} (in factor analysis), \emph{parameter} (in system settings), \emph{column heading} (in spreadsheet), etc. The terms that may cause confusion and should not be used to mean a variable include class, attribute, value, etc.
  \vspace{-1mm}
  \item \textbf{Category} --- is a \emph{state} of a \emph{concept}. A concept can have an infinite number of states (e.g., an integer number). The term is usually used in conjunction with nominal or ordinal variables, which are also referred to as categorical variables. In general, a category is a \emph{valid value} of a variable. The adjective ``valid'' is necessary for a category (e.g., there are two different colors among five apples -- red and green), as a value can also mean an instance (e.g., five apples are of red, red, red, green, green). In a different context, a category or a valid value may be referred to as \emph{options} (in a set of radio buttons), \emph{letter} (information theory), etc. A variable can be defined with an infinite number of categories.
  \vspace{-1mm}
  \item \textbf{Categorization Scheme} --- A categorization scheme typically consists of $N$ \emph{variables} ($N > 0$), and each variable is defined with a set of categories. 
  \vspace{-1mm}
  \item \textbf{Categorization} --- This is a process for labelling a physical object or data object according to the variables and categories defined in a categorization scheme. The process is also referred to as classification.
  \vspace{-1mm}
  \item \textbf{Coding} --- In GT, the term \emph{coding} may imply categorization (e.g., selectively coding), defining categories through abstraction and grouping (e.g., open coding), comparing different optional categorization schemes (e.g., axial coding), or a combination of these in a single process. Its broad definition is likely related to the ambiguity of the word \emph{code}, which may mean a \emph{coding scheme} as well as a \emph{codeword} in a coding scheme.
\end{itemize}

The criteria for optimizing the specification of a categorization scheme (including its variables and categories) include:
\begin{itemize}
  \vspace{-1mm}
  \item The variables are reasonably orthogonal though it is almost impossible to assure their mutual independency.
  \vspace{-1mm}
  \item The categories in a variable must be mutually exclusive.
  \vspace{-1mm}
  \item When a post may feature elements that could potentially belong to different categories, we use word ``main'' in the definition of the variable, and word ``focus on'', ``mainly'', or ``at least'' etc. to define the boundaries between categories.
  \vspace{-1mm}
  \item The categories in a variable must cover all possible objects to be categorized (e.g., all posts).
  \vspace{-1mm}
  \item The boundary between categories should be easily determinable.
  \vspace{-1mm}
  \item Avoid a category “not applicable” as it is subjective to a 3rd party criterion for the reason ``why not applicable'', making this variable difficult to categorize independently.
  \vspace{-1mm}
  \item When a category relies on a human coder's subjective judgment that cannot be easily verified, the coder must write a commentary note.
\end{itemize}

\subsection*{A.2 Computer Coding per Post}
The VisGuides platform can provide each post with information according to the following variables as part of a categorization scheme automatically or algorithmically:

\begin{table}[h]
  \caption{Variables that can be coded by a computer.}
  \label{tab:Computer}
  \centering
  \begin{tabular}{@{}p{2.5cm}@{\hspace{2mm}}p{6cm}@{}}
    \toprule
    \textbf{Variables} & \textbf{Categories}\\
    \midrule
    \textbf{A1: Post ID} & Computer-generated ID numbers\\
    \textbf{A2: Topic} & (1) Cognition, (2) Education, (3) General, (4) Interaction,
                         (5) Medical Visualization, (6) Perception, (7) Site Feedback,
                         (8) Theory, (9) Uncategorized, (10) Visual Design, (11) VR/VE\\
    \textbf{A3: Thread} & Thread ID\\
    \textbf{A4: Thread Title} & Title of the thread or conversation\\
    \textbf{A5: Date} & Date\\
    \textbf{A6: Time} & Time\\
    \textbf{A7: Length-C} & The number of characters (letters, digits, etc.) in the post\\
    \textbf{A8: Length-W} & The number of words in the post\\
    \textbf{A9: Image} & The number of images included in the post\\
    \textbf{A10: TAR} & Thread-Author Relation: I or I1 (initiator), N (non-owner),
                        N2 (The 2nd person jointed thread), N3 (the 3rd person), ...\\
    \textbf{A11: TPR} & Thread-Post Relation: the ordered number of the posts within the thread,
                        1, 2, 3, 4, ...\\
    \bottomrule
  \end{tabular}
\end{table}

\input{sections/table-H1H2.tex}

\subsection*{A.3 Human Coding per Post}
After several iterations of open coding, axial coding, and selective coding, we have converged to six variables for the part of a categorization scheme to support human coding through close-reading. Although it may be possible in the future that some of such coding may potentially be done by a text analysis algorithm, we do not have any reliable algorithm for coding these variables yet. The involvement of humans in the coding process has brought about the most important scholarly benefit for defining the categorization scheme and understanding the relative merits of different ways of defining variables and categories.

During one iteration, we compared our categorization scheme with the Rhetorical Structure Theory (RST) by Mann and Thompson \cite{Mann:1988:TT} and the subsequent ontology by Mitrovi\'{c} et al. \cite{Mitrovic:2017:AC}. We also consulted the categorization scheme by Wachsmuth et al. \cite{Wachsmuth:2017:CL}, before our decisions on six variables, \textbf{H1. Main Purpose}, \textbf{H2. Guideline Mentioning}, \textbf{H3. Evidence}, \textbf{H4. Main Scope}, \textbf{H5. Antithesis}, and \textbf{H6. Conviction}, which are detailed in Tables \ref{tab:H1}-\ref{tab:H6}.

\input{sections/table-H3H4.tex}

\input{sections/table-H5H6.tex}

\subsection*{\textbf{A.4 Our GT Coding Process}}
Our methodology to code information from VisGuides followed the iterations of open coding, axial coding, and selective coding, together with memoing that records all the intermediate stages. 
As a result of applying GT, our coding scheme evolved in four main versions: $S_1$ to $S_4$. 
In this section, we use the following terminology:

\begin{itemize}[noitemsep,leftmargin=*]
    \item a \textbf{post} is an individual question or reply posted in the forum by a person. 
    \item A \textbf{thread} is an initial post plus the temporally-ordered sequence of all the replies, with the same ``(Thread) Topic'' in VisGuides. 
    \item A \textbf{question} is the post starting a thread. 
\end{itemize}

\begin{figure*}[t]
    \centering
    \includegraphics[width=140mm]{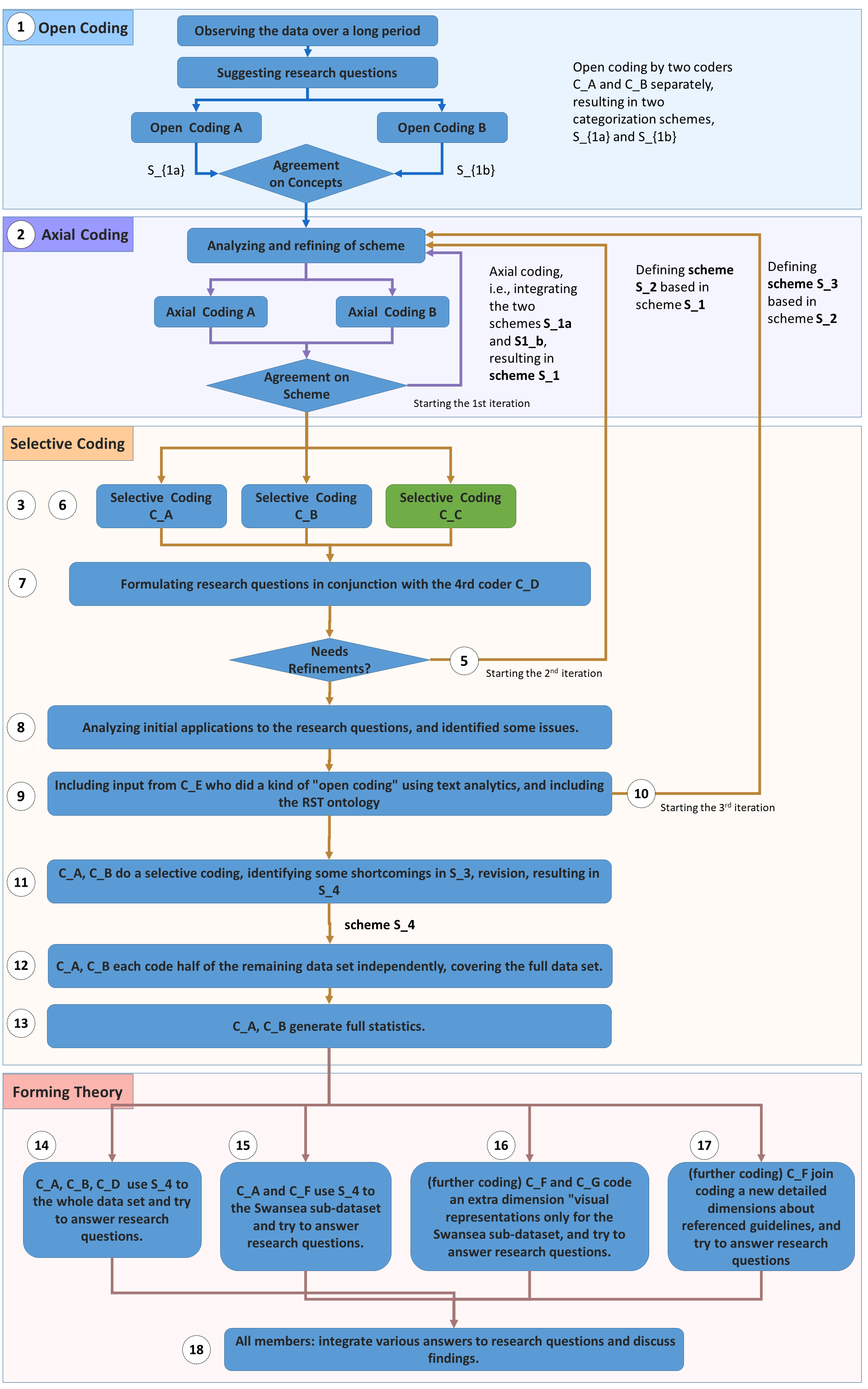}
    \caption{A task workflow chart showing the different steps in our GT coding process. The tasks are grouped in four main stages in GT: open coding, axial coding, selective coding, and theory formation~\cite{dillon2012grounded}.}
    \label{fig:GT-overview}
\end{figure*}

The coding was done mainly by two coders, $C_A$ and $C_B$, both of whom have been managing VisGuides and have read many of the posts before the coding started. Both coders, together with other team members have discussed possible research questions and hypotheses that one may be interested in grounding such questions on the data from VisGuides. Fig.~\ref{fig:GT-overview} shows the steps followed by the team in detail when applying GT.

To construct a categorization scheme, we followed the GT approach to ``orient'' potential variables towards our research questions. 
Example variables included \textit{`What are the questions asking for?'}, \textit{`How specific are the questions asking for advice?'}, and \textit{`Which guidelines are implicitly or explicitly mentioned in the threads'}. From this list of potential variables, we marked out all the information that would not require any human coding, such as \textit{number of replies per post}, \textit{posts per topic}, and dates. 
This left us with some candidate variables to consider in an iterative GT process involving open coding, axial coding, and selective coding.
There are many variables that could be used to characterize each post or thread, including but not limited to, those language variables for coding guideline statements used in \cite{kandogan2016grounded}; variables for coding visualization topics, users, tasks, and applications; and variables about the discourses where guidelines are discussed.
In our coding exercise, our data and our research questions directed us to focus on the variables that are featured in the data and can be used to answer our research questions. 



\begin{enumerate}[noitemsep,leftmargin=*]
  \item \textbf{Staring the first iteration, aiming for $S_1$.} -- The two main coders, $C_A$ and $C_B$ first open-coded the data independently, resulting in two categorization schemes $S_{1a}$ and $S_{1b}$. For this iteration, both coders formulated their variables and categories on a per-\textit{thread} basis, i.e., considering all posts in each thread as a whole. 
  
  \item $C_A$ and $C_B$ then conducted axial coding, integrating the two independent categorization schemes. This resulted in scheme $S_1$, which contained 6 dimensions with 21 categories in total. Additionally, we defined some plain-text categories, characterizing how the discussions in a thread was evolving (e.g., ``single-reply'', ``first reply says the given guideline is too strict, then points to literature for general advice'', etc.), and which guidelines and which topics were mentioned.
  
  \item $C_A$ and $C_B$, with the help of a third coder $C_C$ (an MSc student), then applied $S_1$ to a subset of the data, i.e., selective coding the 10 threads with the most replies. 
  
  \item All three coders $C_A$, $C_B$, $C_C$ then discussed the scheme $S_1$ and the coding results. During the discussion, $C_A$ and $C_B$ agreed that coding on a per-thread basis was too coarse to establish a suitable categorization scheme since threads contained too much and too rich information and thus required the coders to make too many complex decisions.

  \item \textbf{Staring the second iteration, aiming for $S_2$.} -- This time, we aimed for coding on a paragraph-basis, i.e., coding individual paragraphs in each post. Based on the discussion about $S_1$, $C_A$ and $C_B$ refined the dimensions to obtain an $S_2$, which included 5 dimensions (\textit{How is the guideline expressed?} \textit{How suggestive is the statement?}, \textit{Is the statement mentioning any conditions?}, \textit{How is the statement supported?}, and \textit{How generalizable are the statements to other topics in visualization?}). These dimensions included 27 categories in total.

  \item $C_A$ and $C_B$ conducted selective coding independently by applying $S_2$ to a subset of the data. When analyzing the coding results, $C_A$ and $C_B$ found a fair amount of disagreement in the coding results. The coding on a per-paragraph basis again turned out to require too many complex decisions from the coders, including the decision on how to split the posts into meaningful units. Moreover, we found that many paragraphs contain similar information and hence would have preferred to combine them into a higher level unit. In summary, we found coding on a per-paragraph basis too fine-grained. That was an obstacle to continue with the next step, i.e., applying $S_2$ to the whole data set, and we anticipated that the results would not provide adequate support to the stage of ``Forming Theories''. We decided that the best way forward was to code on a per-post basis.
  
  
  
  \item $C_A$ and $C_B$, in conjunction with another researcher $C_D$, formulated research questions \textbf{Q1-Q3} as described in Section~\ref{sec:Introduction}. With the addition of $C_D$, we had established a more focused set of research questions and hypotheses that could be suitably grounded in our data. We discovered several issues related to the completeness and independence of the existing variables.
  
  \item $C_A$ and $C_B$ revised $S_2$, resulting in $S_2'$, in order to make it work with posts, and make it complete, errors free, and independent.
  $C_A$ and $C_B$ conducted another round of selective coding by applying $S_2'$ to the selected subset of the data, and analyzed the coding results to see if they could help answer the research questions, \textbf{Q1-Q3}. $C_A$ and $C_B$ saw the improvement of the new scheme $S_2'$ but also identified some further issues.
  
  \item $C_A$ and $C_B$ received some input from a team member, $C_E$, who had been observing the data using text analytics. As text analysis and visualization can be considered as open coding by the computer, the inputs were informative. $C_E$ also drew attention to the references of Rhetorical Structure Theory (RST).
 
  \item \textbf{Starting the third iteration, aiming for $S_3$.} --
  $C_A$, $C_B$, and $C_D$ conducted axial coding by integrating $S_2$ with some concepts in RST, resulting in scheme $S_3$. $S_3$ that focused on posts as an objective middle-granularity between paragraphs and threads.
   
  \item $C_A$ and $C_B$ selectively coded some data (i.e., the posts in the 10 threads with the most replies), and identified some shortcomings in $S_3$, and a revision resulted in the current categorization scheme $S_4$ discussed in Section \ref{sec:VisGuides} and detailed in Appendix A.3). We agreed on $S_4$ for the categorziation scheme for coding all posts in the data set $D_{20190313}$.
  
  \item $C_A$ and $C_B$ applied $S_4$ to code the full data set, by splitting the remaining data set into two equally sized parts and coding each part independently. 
  
  \item $C_A$ and $C_B$ generated some statistics on the frequency of categories to aid the examination of \textbf{Q1}, shown in Fig~\ref{fig:catstats}. The statistics based on human coding results allowed us to have a first glimpse into the summary overview visualization (as shown in Fig.~\ref{fig:catstats}), confirming some of the patterns that we have already observed during coding.
  
  \item $C_A$, $C_B$, and $C_D$ applied $S_4$ to answer research questions \textbf{Q1-Q2}, and used the computer-coded results to answer research question \textbf{Q3}.

  \item $C_A$ and another team member $C_F$ applied $S_4$ to the Swansea sub-dataset $D_{\text{Swansea2019}}$ and tried to answer research questions \textbf{Q4-Q5} related to the learning activities of Swansea students using their posts at VisGuides.
  
  \item \textbf{Additional Coding.} -- $C_F$ and $C_G$ coded an extra variable of ``visual representations'' for the Swansea sub-dataset to help answer research questions \textbf{Q4-Q5}.
  
  \item \textbf{Additional Coding.} -- another team member $C_H$ joined the coding effort for a new detailed variable about directly- and indirectly-referenced guidelines. This was to help answer research question \textbf{Q2} and the results are reported in Appendix B.
  
  \item We worked collaboratively to integrate the various answers to the research questions posed in the paper. The application of GT allowed us to examine the use of guidelines in different contexts, and identify several findings.
\end{enumerate}

\clearpage

%% file: sections/table-H1H2.tex
\begin{table*}[t]
  \caption{The categories of the variable \textbf{H1: Main Purpose}.}
  \label{tab:H1}
  \centering
  \begin{tabular}{@{}p{2.5cm}@{\hspace{2mm}}p{15cm}@{}}
    \toprule
    \textbf{Variable} & \textbf{Definition and Explanation}\\
    \midrule
    \textbf{H1. Main Purpose} &
    The main purpose of the post (related to ``Purpose'' in RST). The post exhibits more than one purpose, the coder will make a judgment about what the main one is.\\
    \midrule
    \textbf{Category} & \textbf{Definition and Explanation}\\
    \midrule
    \emph{Ask for Advice} &
    A post asks for information about some techniques, guidelines, theories, etc. This includes posts for seeking advice on specific problems, but does not includes posts for seeking clarification from the author of a previous post, nor posts for challenging or critiquing through an expressive question.\\ 
    \emph{Offer Advice} &
    A post that is intended to offer advice, and to pass on knowledge and experience for helping others who are seeking advice. This includes posts that volunteer the author's or someone's guidelines without any prompt, as well as posts that were intended to offer alternative advice that may differ from or in conflict with the advice given by others.\\
    \emph{Discuss Guidelines} &
    A post that focuses on guidelines, it may question the correctness of a guideline or an aspect of a guideline, comment on a guideline, critique guideline, or defend a guideline. This category does not include offering references about guideline, explaining a guideline, or providing advice involving a guideline.\\
    \emph{Seek Clarification} &
    A post seeks a clarification from the author of a previous post that could be asking a question or offering some advice, information, or comment. For example, one may ask ``when you say a colormap, do you mean a continuous or discrete colormap?''\\
    \emph{Add Clarification} &
    A post provides a clarification about some statements in a previous post. To distinguish this from ``ask advice'', ``offer advice'', or ``discuss guideline'', the emphasis is the additionality to the previous post. If the post contains a substantial amount of new information, it is better to use one of the other categories.\\
    \emph{Social Conversation} &
    Acknowledgement, etiquette, or any social conversation, such as ``Thank you.''\\
    \emph{Administration} &
    This is reserved for any administrative posts sent by VisGuides managers.\\
    \bottomrule
  \end{tabular}
\end{table*}

\begin{table*}[t]
  \caption{The categories of the variable \textbf{H2. Guideline Mentioning}.}
  \label{tab:H2}
  \centering
  \begin{tabular}{@{}p{3.8cm}@{\hspace{2mm}}p{13.7cm}@{}}
    \toprule
    \textbf{Variable} & \textbf{Definition and Explanation}\\
    \midrule
    \textbf{H2. Guideline Mentioning} &
    The way in which guidelines are mentioned in the post (``Claim'' and ``Support'' in RST).\\
    \midrule
    \textbf{Category} & \textbf{Definition and Explanation}\\
    \midrule
    \emph{Direct} &
    Mentioning a guideline explicitly, such as according to X, ``one should do this''. If someone proposes a NEW guideline himself or herself, it belongs to this category.\\
    \emph{Indirect} &
    Mentioning a reference to a publication, a blog, but without spelling out a guideline,     e.g., ``there is a fact about color [REF], thus it is better to do this.''\\
    \emph{Implicit} &
    Referring to a reasonably well-known guideline but without mentioning it clearly. For example, ``you may consider maximizing data-ink ratio here''. To select this category, the coder must add a link or reference in the comment section.\\
    \emph{Unclear or Possible} &
    The coder suspects that there might be an implicitly mentioned guideline, but cannot determine any reference or source.\\
    \emph{None} &
    The coder cannot identify any guideline mentioned in the post, and is reasonably sure that no guideline was mentioned in the post.\\
    \bottomrule
  \end{tabular}
\end{table*}

%% file: sections/table-H3H4.tex
\begin{table*}[t]
  \caption{The categories of the variable \textbf{H3. Evidence}.}
  \label{tab:H3}
  \centering
  \begin{tabular}{@{}p{2.5cm}@{\hspace{2mm}}p{15cm}@{}}
    \toprule
    \textbf{Variable} & \textbf{Definition and Explanation}\\
    \midrule
    \textbf{H3. Evidence} &
    The level of details of the evidence the post provides to support something (a technique, a visual design, a guideline, etc.) works, does not work, or needs to be addressed (related to ``Evidence'' and ``Elaboration'' in RST).\\
    \midrule
    \textbf{Category} & \textbf{Definition and Explanation}\\
    \midrule
    \emph{Detailed} &
    One or more examples or case studies are clearly given such as in many posts by Swansea students. There is little imagination required by a reader to understand the evidence.\\
    \emph{Brief} &
    A short description (typically one sentence, one or two references) that a reader has to imagine the actual scenario, or go read the reference. For example, ``this is similar to the technique that worked for bio-visualization.''\\
    \emph{Unclear} &
    The coder suspects that some statements in the post may imply some evidence, but cannot be sure about this.\\
    \emph{None} &
    The coder cannot identify any evidence mentioned in the post, and is reasonably certain that the post does not include any evidence.\\
    \bottomrule
  \end{tabular}
\end{table*}

\begin{table*}[t]
  \caption{The categories of the variable \textbf{H4. Main Scope}.}
  \label{tab:H4}
  \centering
  \begin{tabular}{@{}p{2.5cm}@{\hspace{2mm}}p{15cm}@{}}
    \toprule
    \textbf{Variable} & \textbf{Definition and Explanation}\\
    \midrule
    \textbf{H4. Main Scope} &
    The scope that the post covers or intends to cover (related to ``Circumstance'' and ``Condition'' in RST). When a post has different types of coverage, the coder will make a judgment as the main coverage intended for this post. For example, if a post focuses on advising a specific design of an interaction, suggesting ``you should use `radio buttons' for this interaction, though I prefer check boxes in general,'' one may consider this to be a case-based scenario.\\
    \midrule
    \textbf{Category} & \textbf{Definition and Explanation}\\
    \midrule
    \emph{Case-based} &
    The post focuses on one or more individual cases, such as the problems related to a specific data set or visual design in most posts by Swansea students.\\
    \emph{Conditional} &
    This post focuses on a good number of cases that may fall under certain conditions. Here condition is defined as a smaller scope than context. Examples include ``can parallel coordinates deal with more than 100 axes?'', ``how can one design a multivariate bar charts without using colors?'', and ``this guideline does work in most situations, but not in the following conditions.''\\
    \emph{Contextual} &
    This post focuses on a specific context that can easily cover many example cases. The contexts may be defined at a relatively high-level, such as topics (e.g., interaction, color), visual representations (bar charts), users (e.g., domain experts), tasks (navigation), applications (e.g., sports), and so on.\\
    \emph{General} &
    The post focuses on a very general discussion about the field of visualization or beyond such as computer science, data science.\\
    \emph{Unclear} &
    The coder detects some focal topics in the post, but cannot be certain about its scope.\\
    \emph{No Focal Topic} &
    The coder is reasonably sure that this post does not have a focal topic. This includes those posts for acknowledgement, and social conversation.\\
    \bottomrule
  \end{tabular}
\end{table*}

%% file: sections/table-H5H6.tex
\begin{table*}[t]
  \caption{The categories of the variable \textbf{H5. Antithesis}.}
  \label{tab:H5}
  \centering
  \begin{tabular}{@{}p{2.5cm}@{\hspace{2mm}}p{15cm}@{}}
    \toprule
    \textbf{Variable} & \textbf{Definition and Explanation}\\
    \midrule
    \textbf{H5. Antithesis} &
    The Level of Agreement with previous posts, excluding any future posts (related to ``Antithesis'' in RST). The categories are defined in relation with all previous posts in the thread. It is not about agreement with a particular guideline, opinion, or advice, which will not be easy to code without specifying which guideline, opinion, or advice.\\
    \midrule
    \textbf{Category} & \textbf{Definition and Explanation}\\
    \midrule
    \emph{Neutral} &
    The post does not show any opinion that could possibly be in disagreement with previous post. This includes typical posts for seeking information or advice. However, it does not include posts that challenge or question existing guidelines.\\
    \emph{So Far So Good} &
    The post has an opinion, poses a challenge to a guideline, or offer an advice, but it has not yet in disagreement with previous posts (implying all previous posts are in agreement). This includes the first post for opening a thread with a comment or question that challenges a guideline. \\
    \emph{Objection} &
    The post has at least one opinion that disagrees with some opinions in the previous posts. This includes posts for defending one's own opinions after being criticized.\\
    \emph{Unclear} &
    The coder detects some opinions in the post, but cannot be certain whether they are in disagreement with any previous posts.\\
    \bottomrule
  \end{tabular}
\end{table*}

\begin{table*}[t]
  \caption{The categories of the variable \textbf{H6. Conviction}.}
  \label{tab:H6}
  \centering
  \begin{tabular}{@{}p{2.5cm}@{\hspace{2mm}}p{15cm}@{}}
    \toprule
    \textbf{Variable} & \textbf{Definition and Explanation}\\
    \midrule
    \textbf{H6. Conviction} &
    The Level of Conviction shown by the arguments in the post (related to ``Rhetorical'' and ``Concession'' in RST).\\
    \midrule
    \textbf{Category} & \textbf{Definition and Explanation}\\
    \midrule
    \emph{Neutral} &
    The post does not show any opinion. This includes typical posts for seeking information or advice. However, it does not include posts that change or question existing guidelines.\\
    \emph{Doubtful} &
    The post expresses some doubts about a guideline or some opinions in the previous posts (e.g., through some mild questions) but does not explicitly give an opinion ``for'' or ``against''.\\
    \emph{Suggestive} &
    The post has at least one opinion, and all opinions are expressed in an uncertain way, with words such as ``perhaps'', ``possibly'', etc.\\
    \emph{Affirmative} &
    The post has at least one affirmative opinion, with words such as ``must'', ``should'', ``cannot be'', ``should never'', etc.\\
    \emph{Unclear} &
    The coder cannot determine if the post contains any opinion, e.g., the English expression in the post is difficult to understand.\\
    \bottomrule
  \end{tabular}
\end{table*}

%% file: sections/appendix-B.tex
\section*{\textbf{B. Mentioned Visualization Guidelines}}
\label{app:Guidelines}

This section lists all guidelines that are directly and indirectly mentioned in the posts between September 27, 2017 and March 13, 2019.

\subsection*{\textbf{B.1 Directly Mentioned Visualization Guidelines}}
\label{app:Direct}

The following guidelines have been directly mentioned in the discourses at VisGuides, and they are listed in the alphabetic order of the first keyword highlighted in bold. For guidelines with the same first keyword, the ordering is based on the second keyword, and so forth.   

\begin{itemize}[noitemsep,leftmargin=*]
  \item Do not use the \textbf{blow apart effects}. \cite{Duke:2019:web}
  
  \item The \textbf{Correspondence} Principle---The same data should produce the same visualization~\cite{Kindlmann:2014:TVCG}.
  
  \item Good visualization should maximize the \textbf{data-ink ratio}~\cite{tufte2001visual}.
  
  \item Facilitate \textbf{depth perception} for \textbf{3D} visualizations~\cite{borgo2013glyph}.
  
  \item \textbf{Feature-driven animation} vs. \textbf{interactive systems}:``Start with an overview of entire duration. End with a focused view in relevant time duration. Include all relevant data attributes in at least one segment of the animation sequence~\cite{Yu:2017:EI}.
  
  \item Shneiderman's  Visual \textbf{Information-Seeking Mantra}: Overview first, zoom and filter, details on demand~\cite{Shneiderman:1996:VL}.
  
  \item Use \textbf{interaction} in visualization sparsely and cautiously~ \cite{Groller:2008:Dagstuhl}.
  
  \item \textbf{Lie factor} guideline~\cite{tufte2001visual}: The representation of numbers, as physically measured on the surface of the graphic itself, should be directly proportional to the quantities represented.
  
  \item Don't use \textbf{more than six colors together}~\cite{Sayers:2019:web}.
  
  \item The \textbf{rainbow color} map is considered harmful~\cite{Rogowitz:1998:S,Borland:2007}

  \item \textbf{Reduce clutter} in parallel coordinates by reordering axes~\cite{Makwana:2016:IJCSIT}. 

  \item Elmqvist's Rule \#7: Don't \textbf{replicate} the \textbf{real world}~\cite{Elmqvist:2017:blog}.

  \item Few's Rule \#9: Avoid using \textbf{visual effects} in \textbf{graphs}~\cite{Few:2008:web}.
 
 \item \textbf{Visual variables} for visualizations in \textbf{VR}~\cite{Mackinlay:1986:TOG}
\end{itemize}

\subsection*{\textbf{B.2 Indirectly Mentioned Visualization Guidelines}}
\label{app:Indirect}

The following references to books, research papers, blogs, video, etc. have been been mentioned in the discussions on various topics, but there was no guideline mentioned explicitly in relation to these references. We consider that it is highly likely that most of these references contain some visualization guidelines. The references are listed according to firstly the topic categories and then the threads where the corresponding posts appeared. The topic categories are ordered alphabetically, and the thread headings are ordered alphabetically according to its keywords.

\paragraph{\textbf{Perception}}
\begin{itemize}[noitemsep]
    \item The \textbf{rainbow color} map is considered harmful~\cite{MPL:2015:web, Rhyne:2016:book}.
\end{itemize}

\paragraph{\textbf{Theory}}
\begin{itemize}[noitemsep]
  \item \textbf{Data-Ink Ratio} Principle, How to use it?~\cite{Strunk:2009:book,Behrisch:2019:CGF}.
\end{itemize}

\paragraph{\textbf{Visual Design}}
\begin{itemize}[noitemsep]
  \item \textbf{Clutter reduction} technique in \textbf{parallel coordinates}: \cite{Heinrich:2013:STAR}.
  \item Visual \textbf{Information-Seeking Mantra}: \cite{vanHam:2009:TVCG,Chen:2016:book,Luciani:2019:TVCG}.
  \item \textbf{Lie factor} guideline: \cite{Meihoefer:1973:Cartographica, Cleveland:1984:JASA}.
  \item Elmqvist's Rule \#7: Don't \textbf{replicate} the \textbf{real world}: \cite{Bach:2018:TVCG, Besancon:2017:CHI}.
  \item \textbf{Visual variables} for visualizations in \textbf{VR}: \cite{Chen:2014:BookChapter}
  \item \textbf{Visualizing time} on a \textbf{geospatial map}: \cite{Weber:2001:InfoVis, VanWijk:1999:InfoVis} 
\end{itemize}

%% file: sections/appendix-C.tex
\begin{figure*}[t]
    \centering
    \includegraphics[width=180mm]{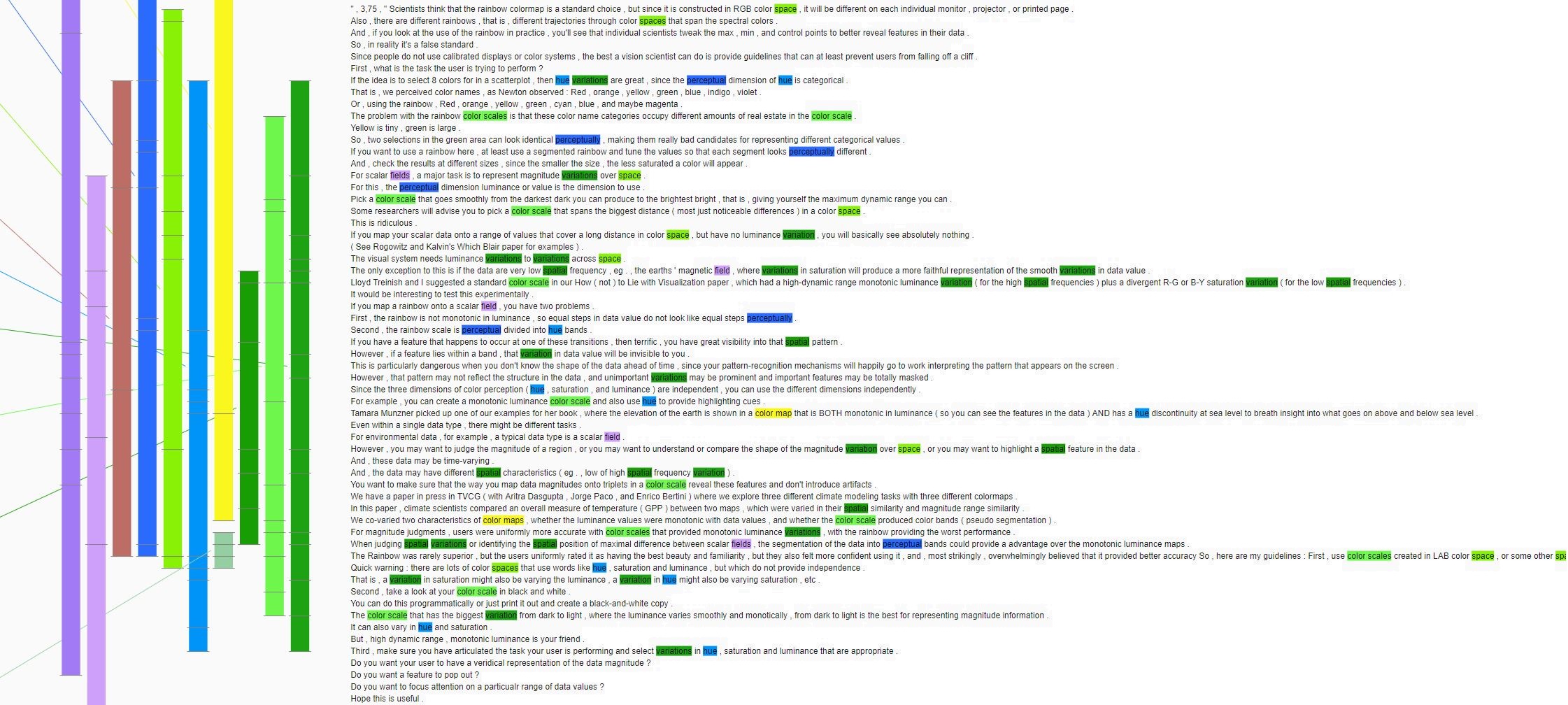}
    \caption{A closely zoomed-in view of a Lexical Episode Plot [30], revealing the interesting text areas for close-reading.}
    \label{fig:LEX-text}
\end{figure*}

\section*{\textbf{C. Further Text Visualization Images}}
\label{app:TextAnalytics}
This appendix contains additional visualization images that would br too large for inclusion in Section \ref{sec:TextAnalytics}.

\begin{figure*}[t]
    \centering
    \includegraphics[width=180mm, trim={2.3cm 0 1.6cm 0}, clip]{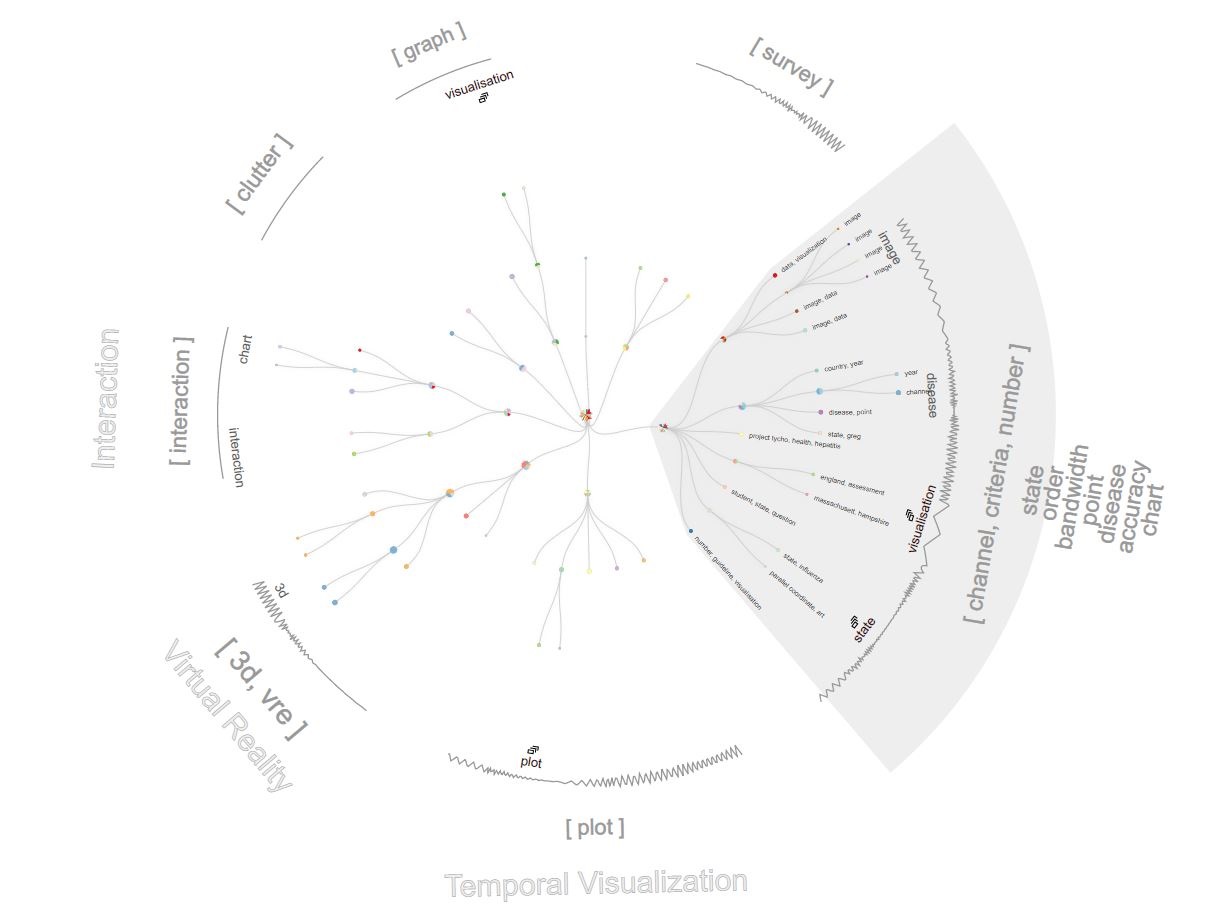}
    \caption{An intermediate result of the  Topic Tree~\cite{el2018visual}, which \textit{groups all semantically-related utterances} (as leaf nodes) in a hierarchical tree structure with inner-nodes constituting topics and sub-topics. The \textit{outer border} of the tree depicts the \textit{top-descriptors} of each topic branch and the \textit{uncertainty of the respective utterance assignment} to that topic. Hovering over a topic segment reveals the top descriptive keywords for that topic.}
    \label{fig:LEP-overview}
\end{figure*}

From Fig. \ref{fig:LEP-overview}, we can observe more potential keywords (e.g., \emph{interaction, parallel coordinates, 3D, bubble, etc.} than the zoomed-in view in Fig. \ref{fig:LEP}. Fig. \ref{fig:LEX-text} shows a closely zoomed-in view, enabling users to conduct close-reading of the text that are associated with the identified keywords.

\begin{figure*}[t]
    \centering
    \includegraphics[width=140mm]{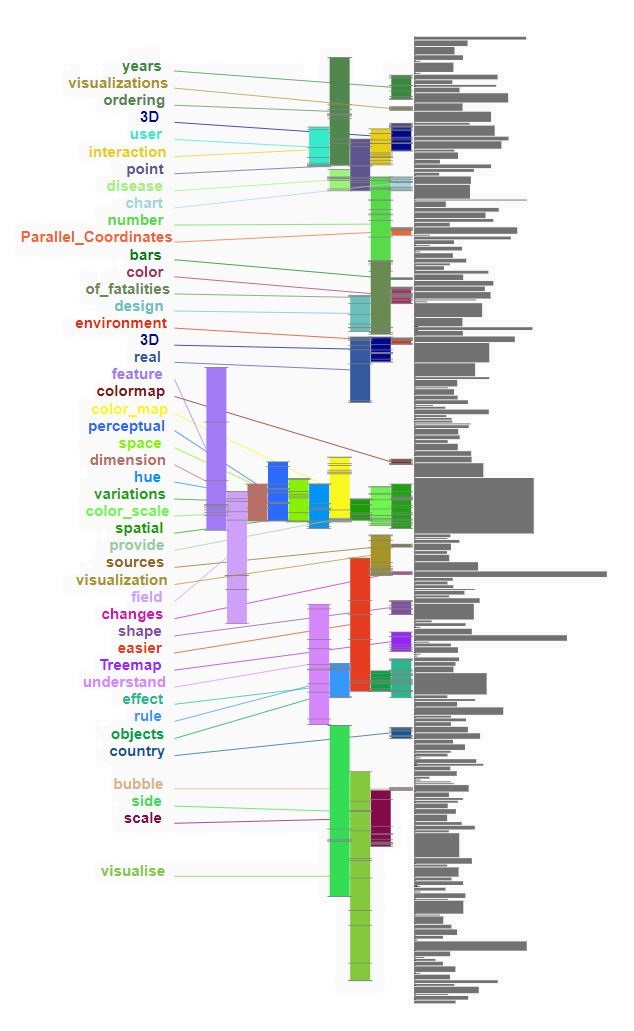}
    \caption{A zoomed-out view of a Lexical Episode Plot [30] showing the evolution of the discussions over time. It automatically detects compact chains of $n$-grams and highlights them beside the complete text as an overview. One can interactively zoom-in to the interesting text areas for close-reading.}
    \label{fig:LEX-overview}
\end{figure*}

\begin{figure*}[t]
    \centering
    \includegraphics[width=180mm]{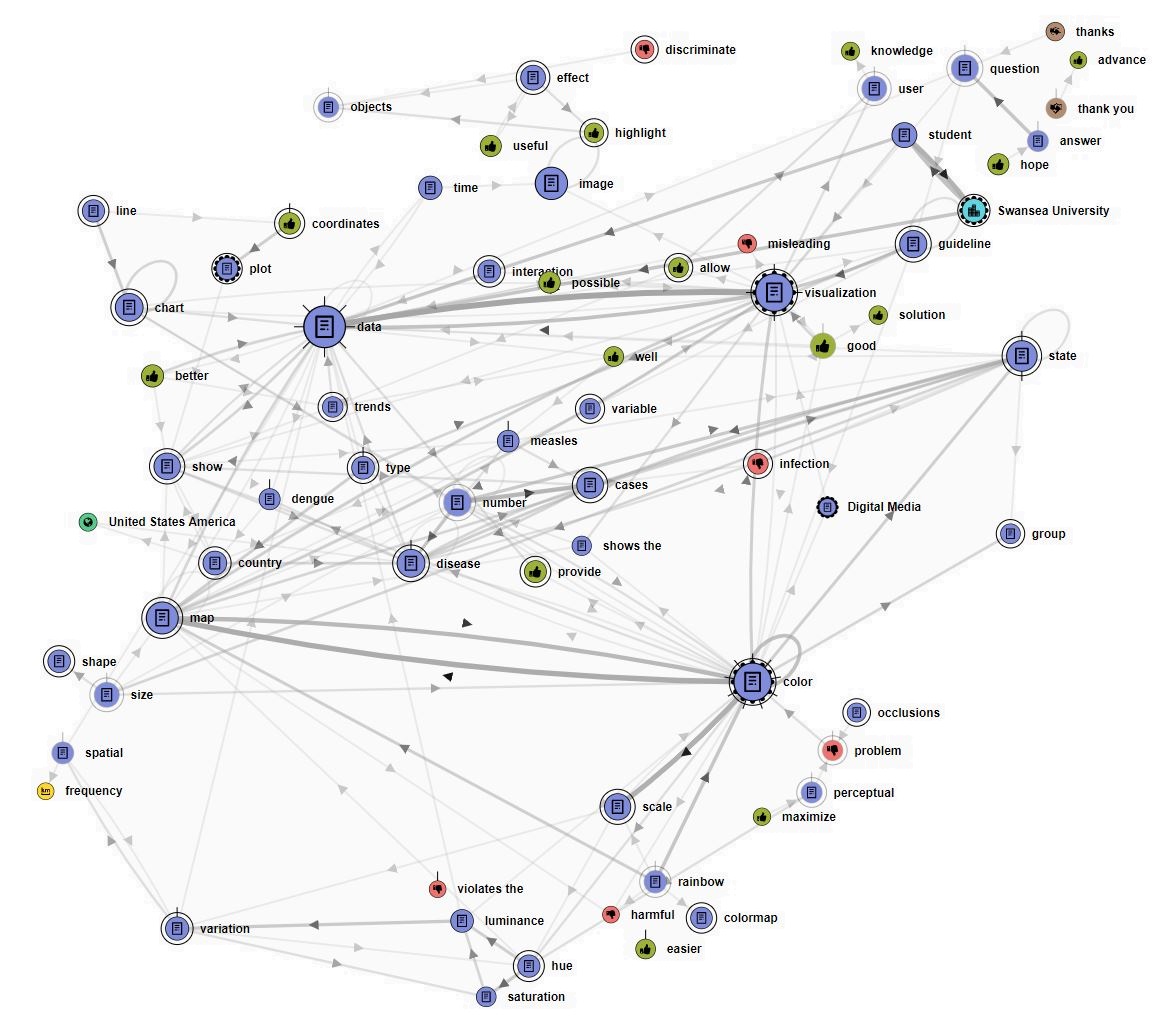}
    \caption{The Named Entity Graph~\cite{El-Assady2017} depicts the relationship between the most \textit{relevant} entity pairs in a corpus, i.e., entities occurring more often together than an expected threshold. In this graph instance,  the node ``color'' is very prominent, as the use of rainbow colormaps has been extensively discussed in the underlying text.}
    \label{fig:NEREx}
\end{figure*}

\begin{figure*}
    \centering
    \includegraphics[width=170mm]{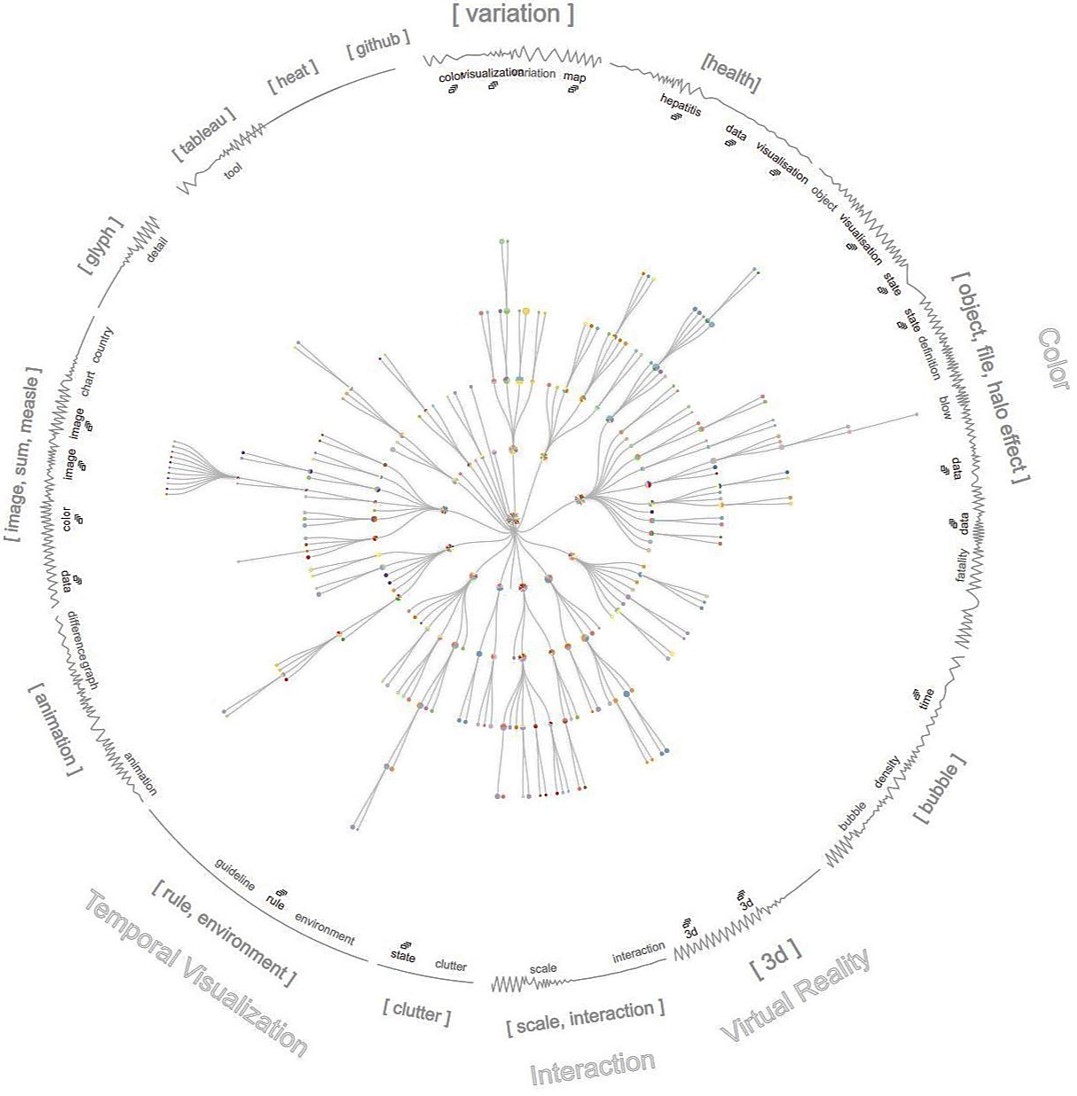}
    \caption{The Topic Tree~\cite{el2018visual} \textit{groups all semantically-related utterances} (as leaf nodes) in a hierarchical tree structure with inner-nodes constituting topics and sub-topics. The \textit{outer border} of the tree depicts the \textit{top-descriptors} of each topic branch and the \textit{uncertainty of the respective utterance assignment} to that topic. }
    \label{fig:topics}
\end{figure*}

%% file: sections/appendix-D.tex
\section*{\textbf{D. Grounded Theory and VisGuides Data}}
\label{app:OriginalGT}
\subsection*{\textbf{D.1 Formal Descriptions of Ground Theory Methods}}
\emph{Grounded theory} was proposed in the context of the social sciences \cite{Glaser:1967:book}. Consider a postulated theory $T$ (e.g., about social behaviors), which may be in the form of a categorization scheme, a taxonomy, an ontology, a causal relation, a conceptual model, or a numerical model. The initial theory $T_1$ may be formulated based on a data set $D_1$ (e.g., a collection of news reports on social behaviors). We use $D_1 \hookrightarrow T_1$ to denote the grounded process for deriving $T_1$ from $D_1$.

Grounded theory instigates that the theory $T$ needs to be grounded in the data space $\mathbb{D}$ where $T_1$ emerged, by constantly and continuously sampling the data space, analyzing the data captured, and refining the theory until it reaches theoretical saturation $T_\text{sat}$. In other words, over a long period of time, the sampling collects new data sets $D_2, D_3, \ldots, D_t, \ldots, D_n \in \mathbb{D}$, and the analysis evolves the theory as $T_2, T_3, \ldots, T_t, \ldots, T_\text{sat}$.
At each iteration $t$ where a new data set $D_t$ is obtained, grounded theory outlines a number of methods for obtaining $T_t$. As listed in Table \ref{tab:GroundedTheory} that was compiled based on \cite{Willig:2013:book}, these include:
\begin{itemize}
    \vspace{-1mm}%
    \item \textbf{Categorization}: deriving a refined theory based on the combined data collection as $\bigcup_{i=1}^t D_i \hookrightarrow T_t$, with aid of other methods below. Here $\bigcup$ denotes the union of data sets.
    \vspace{-1mm}%
    \item \textbf{Coding}: deriving candidates of $T_i$ through close reading of the data set $D_i$. For example, using \emph{open coding}, different people,  $a, b, c, \ldots$, may derive several candidate theories independently:
    \vspace{-1mm}
    \begin{equation*}
      D_i \stackrel{a}{\hookrightarrow} T_{i,a}, \quad
      D_i \stackrel{b}{\hookrightarrow} T_{i,b}, \quad
      D_i \stackrel{c}{\hookrightarrow} T_{i,b}, \quad \ldots   
      \vspace{-1mm}
    \end{equation*}
    \emph{Axial coding} enables the identification of the relationships among the candidate theories, while \emph{selective coding} applies the existing and new candidate theories to different data sets, e.g., $\Xi(T_{i-1}, D_i), \Xi(T_{i,a}, D_{i-1}),$ etc., where we write the application of a theory $T$ using a data set $D$ as a function $\Xi(T, D)$.
    \vspace{-1mm}%
    \item \textbf{Comparative Analysis}: comparing the existing and new candidate theories, for example, by evaluating the applications of these theories to all collected data, $\Xi(T_{i-1}, \bigcup_{i=1}^t D_i), \Xi(T_{i,a}, \bigcup_{i=1}^t D_i),$ etc. or by integrating two or more theories into a new candidate theory and then continuing the comparative analysis. 
    \vspace{-1mm}%
    \item \textbf{Negative Cases Analysis}: identifying negative cases during the coding and comparative analysis, gaining a deep understanding of these cases, and if appropriate, refining the proposed theories.
    \vspace{-1mm}%
    \item \textbf{Memoing}: recording the ideas and actions that lead from the data $D_1, D_2, \ldots, D_i$ to various intermediate theories (e.g., $T_{i,a}, T_{i,b},$ etc.) and eventually to $T_i$.
\end{itemize}

The aforementioned methods embody the three principles of grounded theory, namely:
\begin{itemize}
    \vspace{-1mm}%
    \item \textbf{Theoretical sensitivity}: a theory should be derived from data, validated by data, and remain sensitive to the data \cite{Glaser:2004:FQSR}.
    \vspace{-1mm}%
    \item \textbf{Theoretical sampling}: data collection should be oriented towards the generation, evaluation, and refinement of a theory or theories \cite{Breckenridge:2009:GTR}.
    \vspace{-1mm}%
    \item \textbf{Theoretical saturation}: the process of grounding a theory on data should be constant and continuous until all constructs of a theory is fully represented by the data \cite{Saunders:2018:QQ}.
\end{itemize}

\subsection*{\textbf{D.2 Grounding Visualization Guidelines on Data.}}
From the perspective of GT, visualization guidelines are theories that should be grounded on data, since almost all these guidelines were proposed based on the originators' observation and experience, or on the results of empirical studies.

An ideally-formulated visualization guideline $T$ may be expressed as a \emph{construct} ``doing action $\mathbf{A}$ causes outcome $\mathbf{B}$ under the condition $\mathbf{C}$.''
For example, Correll and Gleicher proposed a guideline: ``in bar charts ($\mathbf{C}$), error bars ($\mathbf{A}$) considered harmful ($\mathbf{B}$)'' \cite{Correll:TVCG:2014}.
Many guidelines are expressed with a construct ``do $\mathbf{A}$'' or ``don't do $\mathbf{A}$.''
The former implies a positive outcome $\mathbf{B}$ (e.g., ``make sure all data adds up to 100\%'' \cite{Kosara:2010:web}), while the latter implies a negative outcome $\mathbf{B}$ (e.g., ``do not use blow-apart effects'' \cite{Duke:2019:web}).

Many guidelines do not explicitly specify any condition $\mathbf{C}$, e.g.,
\begin{center}
\vspace{-2mm}
``Rainbow color map is considered harmful'' \cite{Rogowitz:1998:S,Borland:2007}.\\
``Good visualizations should maximize data-ink ratio'' \cite{tufte2001visual}.\\
``Don't use more than six colours together'' \cite{Sayers:2019:web}.\\
``Pie charts are bad and 3D pie charts are very bad'' \cite{Fenton:2009:web}.   
\end{center}
\vspace{-2mm}
\noindent The omission of $\mathbf{C}$ may suggest the general applicability or a deficit of theoretical sensitivity. There is little doubt that in some conditions $\mathbf{C}_1, \mathbf{C}_2, \ldots$, such guidelines are certainly correct. However, we often do not know what these conditions are, suggesting a need for further theoretical sampling. 

In addition to considering visualization guidelines as individual theories, we can anticipate the possibility of formulating meta-theories that model different aspects of the corpus of visualization guidelines, e.g., their categorization \cite{kandogan2016grounded} and their lifecycle \cite{engelke2018visupply}.
According to grounded theory, such meta-theories should also be grounded on the relevant data, which may include different expressions of the guidelines, the attributes of each expression (e.g., originator, location, period, context, etc.), and documentation about their formulation, discussion, curation, sharing, application, evaluation, refinement, and so on. 

\subsection*{\textbf{D.3 VisGuides: Enabling Grounded Theory Methods}}
VisGuides (\url{visguides.org}) is an online discussion forum dedicated to visualization guidelines \cite{Diehl:2018:EuroVis}.
It allows registered users to propose and recommend guidelines, pose questions and offer advice, share positive and negative experience about certain guidelines, report and reason about successes, failures, and conflicts of guidelines, and suggest ways to refine guidelines.
It is a digital platform that features both the ``theories'' (i.e., visualization guidelines) to be developed, and the ``data'' (i.e., discourses) that can be used to generate, evaluate, and refine the theories.
It can support a number of grounded theory methods.

Let $\mathbb{D}_{\text{VisGuides}}$ be the data space of all possible posts at VisGuides. At a point of time $t$, $D_t \in \mathbb{D}_{\text{VisGuides}}$ is the set of posts collected until that time. Hence, VisGuides provides an evolving data set, $D_1 \subseteq D_2, \ldots, D_{t-1} \subseteq D_t$, which is ideal for supporting the principle of theoretical sampling. For a specific context $\Theta$ (e.g., about a datatype, a visual design, a task, an application domain, etc.), one can extract all relevant posts from $D_t$ such that $G$ can be grounded on $D_{t,\Theta} \subset D_t$.
For a specific guideline $G$, this is a special case of context where $\Theta = G$, VisGuides may not have collected sufficient data for grounding $G$ on $D_{t,G}$. A researcher may initiate new discussion about $G$ on VisGuides, or supplement $D_{t,G}$ with data, $E \notin \mathbb{D}_{\text{VisGuides}}$, that is collected outside of VisGuides.

In the following four sections, we will describe a series of research activities that were inspired and guided by the grounded theory methodology. While we employed traditional grounded theory processes, we also utilized our own expertise in computer-assisted data analysis and visualization to complement the traditional processes. These activities are summarized below, where the same tags used in Table \ref{tab:GroundedTheory}, such as $\S_{\ref{sec:VisGuides}} \backsim \S_{\ref{sec:EmpiricalStudies}}$ and $\S_{\text{A}} \backsim \S_{\text{A}}$, relate the GT processes to individual sections and appendices.

\begin{itemize}
    \item \textbf{Traditional Categorization and Coding ($\S_{\ref{sec:VisGuides}}$, $\S_{\ref{sec:Teaching}}$, $\S_{\text{A}}$, $\S_{\text{B}}$).} 
   We close-read all 248 posts in the data set $D_{20190313}$, carried out open and axial coding in order to derive categorization schemes as candidate theories. We used selective coding and statistical graphics to evaluate these categorization schemes. We then proposed an integrated categorization scheme that allowed us to obtain several findings.
    \item \textbf{Text Analysis and Visualization (Section {\ref{sec:TextAnalytics}}, $\S_{\ref{sec:TextAnalytics}}$, $\S_{\text{C}}$).}
    In conjunction with the traditional approach, we applied several text analysis and text visualization techniques to $D_{20190313}$, which allowed us to observe many patterns that could not be derived from close reading easily.
    \item \textbf{Visualization Coursework (Section \ref{sec:Teaching}, $\S_{\ref{sec:Teaching}}$).} With hundreds of guidelines available in the literature \cite{kandogan2016grounded}, it would take a very long time for an individual guideline to receive a good number of hits in sampling. When VisGuides was used to support a piece coursework, we took the opportunity to define a specific context in which the number of applicable guidelines is much smaller and the sampling can be much more intensive. We denote this subset of data as $D_{\text{Swansea2019}} \subset D_{20190313}$. The close-reading and coding of this $D_{\text{Swansea2019}}$ revealed further findings.
    \item \textbf{Empirical Studies (Section \ref{sec:EmpiricalStudies}, $\S_{\ref{sec:EmpiricalStudies}}$).}
    When our initial examination on two specific guidelines indicated a lack of data, we followed the GT methodology to collect more data outside VisGuides. We conducted two empirical studies that resulted in two additional data sets $E_{\text{KCL2019}}$ and $E_{\text{Konstanz2019}}$. These data sets provided the two guidelines with supporting evidence.      
\end{itemize}

\clearpage

%% file: sections/appendix-E.tex
\section*{\textbf{E. Further Details of Empirical Studies}}

\subsection*{\textbf{E.1 A Study Inspired by ``Action without Interaction''}}
This section gives some further details about the apparatus used and the procedure.

\begin{figure}[th!]
\includegraphics[width=\columnwidth]{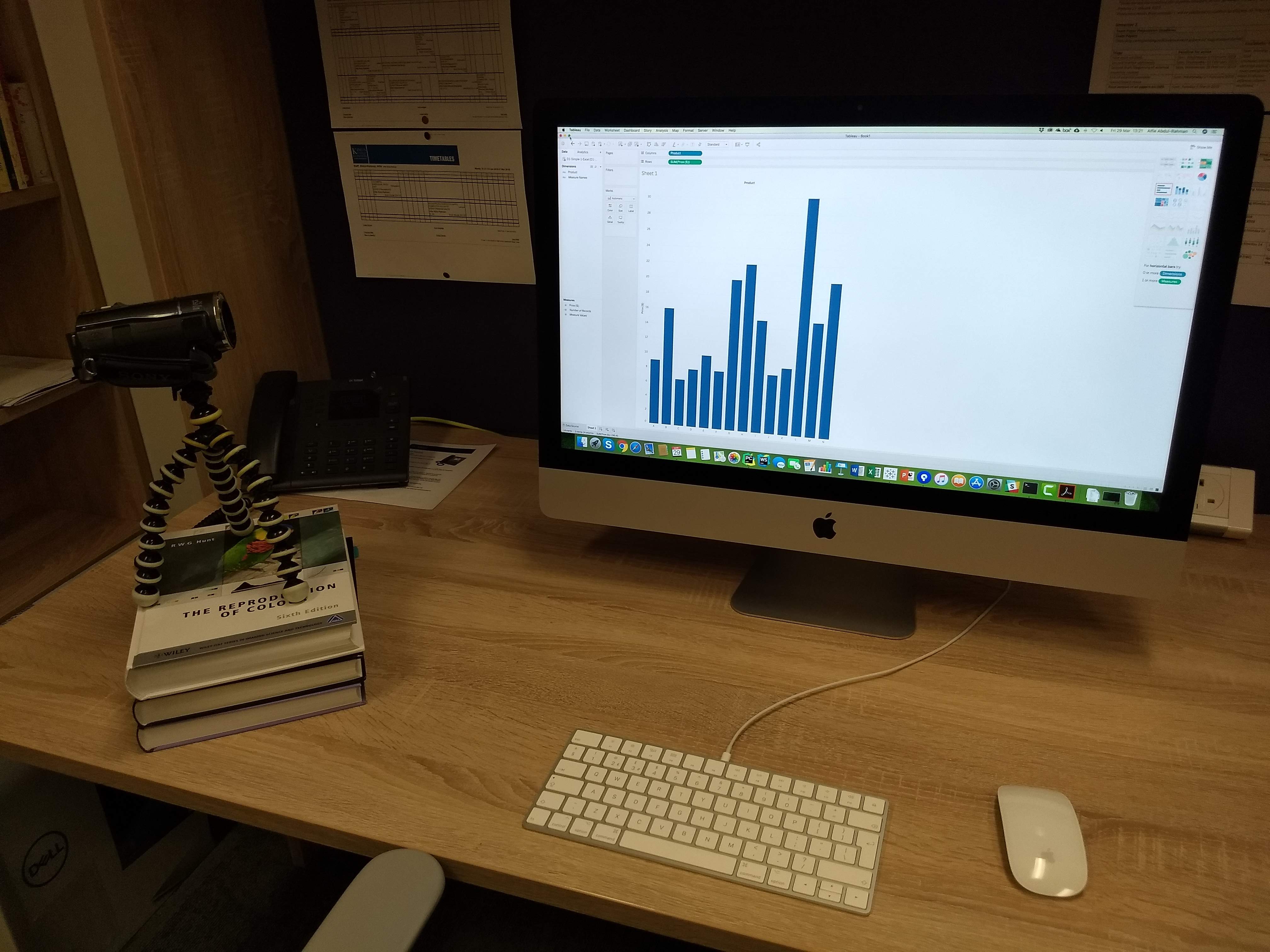}
  \centering
  \caption{Apparatus used in the empirical study on the cost of interaction.}
 \label{fig:KCL-apparatus}
\end{figure}

\paragraph{\textbf{Apparatus.}}
The \emph{apparatus} used for this study includes MS-Excel for Mac (version 16.23) and Tableau Desktop Professional Edition (version 2018.3.3). Both visualization tools were run on macOS High Sierra (version 10.13.6), iMac (Retina 5K, 2017), 3.4 GHz Intel Core i5, 32GB 2400 MHz DDR4, and Radeon Pro 570 4096 MB. Each participant was required to interact with the visualization tool using a keyboard and a mouse at the desk as shown in Fig. \ref{fig:KCL-apparatus}. All trials were recorded by both Camtasia and a camera.

\paragraph{\textbf{Procedure.}} We arranged for the participants to take part in the experiment individually. The time taken to complete the experiment was approximately 10-25 minutes, excluding the process for obtaining consent and the pre-study explanation (5 minutes).

Following the protocol of the university an information sheet was first given to the participant during the recruitment process. Before the pre-study explanation, a consent form was distributed for the participant to read and sign. Once the participant has agreed to take part in the study, the experimenter gave a brief explanation to the participant. The explanation included the tasks involved, the access to the help-sheet, and the sequence of the trials. The participants were informed that they could take as much time as they needed in the experiment.

Following the pre-study explanation, each participant was then given a unique anonymous ID, and a demographic form to be completed for information such as gender, age group, occupation, and familiarity with the two visualization tools tested.
For each participant, the experiment consisted of four trials, either [EXL-Sa, TBL-Sb, TBL-Ca, EXL-Cb] or [TBL-Sa, EXL-Sb, EXL-Ca, TBL-Cb]. In each trial, the data set was pre-loaded onto the visualization tools. Participants were given written specification of the tasks together with examples of the bar charts to be created. After an initial attempt of 1 minute for tasks associated with data sets Sa and Sb and 2 minutes for tasks with data sets Ca and Cb, participants were allowed to make use of a help-sheet to complete the task. At the end of the second trial, participants ware presented with two questions (for T1 and T2 with Sa and Sb) about the easiness of using the two visualization tools. They are asked to rank the relative easiness using the Likert scale. Similarly at the end of the fourth trial, participants were presented with three questions (for T1 and T2 with Ca and Cb, and for the overall impression).

\subsection*{\textbf{E.2 A Study Inspired by ``Don't Replicate the Real World''}}

This section gives some further details about the apparatus used and the procedure.

\paragraph{\textbf{Apparatus.}}
As shown in Fig. \ref{fig:teaser}(c), the main \emph{apparatus} used is a HTC Vive Pro VR headset (GTX1080Ti, SSD, and 16GB RAM). The experiment was carried out in a $7m \times 7m$ physical laboratory, where each participant performed their tasks in the middle of the room while the experimenter sat a few meters away by the computer that controls the virtual environment.
As shown in Fig. \ref{fig:Konstanz-stimuli}, the plain 3D environment contains only a cube on a plane. The top of the cube is the reference plane for the two height fields. The  visually-realistic VE has a $1m^2$ table in the middle of an office environment (walkable area: $5m \times 5m$). The top of the table is the reference plane for the two height fields. 
The virtual office itself is of $7m \times 20m$ including all spaces for the background objects.


\paragraph{\textbf{Procedure.}}
All participants first attended an introduction session. With aid of presentation slides, we described concept of height fields and their typical visual representations, such as heat-maps and 3D surfaces (height-maps). Using examples of real world data (e.g., geographical maps with bi-variate distributions of crime rates and arrest rates, participants became accustomed to the visual observation of height fields, the interpretation of peaks and valleys, and the recognition of similar geometric features between two height fields. They were made aware of the tasks to be performed during the experiment, i.e., identifying common peaks.

The participants would then undergo a training session for working with the VR equipment, including interactive controls for navigation and command selection. Each participant was asked to complete 10 interactive tasks, and was required to repeat the exercise until all tasks were completed without errors.

Participants are randomly divided into two groups. One group began with a block of 15 trials with the plain 3D environment, and then continued with a block of 15 trials with the office-based VE. Another group began with the office-based VE and continued to the plain 3D environment. The stimuli data sets (i.e., the height fields) used in the 30 trials were randomized.